\documentclass[preprint]{aastex}

\usepackage{hyperref}
\usepackage{epsfig}
\usepackage{natbib}
\usepackage{graphicx}                
\usepackage{lscape}
\usepackage{verbatim}              
\usepackage{amsmath,amsthm,amsfonts,amsopn,amssymb} 
\usepackage{longtable}
\usepackage{pdflscape}
\usepackage{afterpage}
\usepackage{rotating}					
\usepackage{ulem}
\usepackage{epstopdf}  
\usepackage{multirow}
\usepackage{color}
\usepackage{graphicx}
\usepackage[left]{lineno}
\newcommand{\rhk}{\mbox{$\log R^\prime_{\rm HK}$}}

\received{}
\accepted{}

\slugcomment{to appear in ApJ}
\shorttitle{Planet Hunters 5}
\shortauthors{Wang}

\begin{document}
\title{Planet Hunters. V. A Confirmed Jupiter-Size Planet in the Habitable Zone and
42 Planet Candidates from the $Kepler$ Archive Data\altaffilmark{1}}
\author{
Ji Wang\altaffilmark{2},
Debra A. Fischer\altaffilmark{2},
Thomas Barclay\altaffilmark{3,4},
Tabetha S. Boyajian\altaffilmark{2},
Justin R. Crepp\altaffilmark{5},
Megan E. Schwamb\altaffilmark{6,7},
Chris Lintott\altaffilmark{8,9},
Kian J. Jek\altaffilmark{10},
Arfon M. Smith\altaffilmark{9},
Michael Parrish\altaffilmark{9},
Kevin Schawinski \altaffilmark{11},
Joseph R. Schmitt\altaffilmark{2},
Matthew J. Giguere\altaffilmark{2},
John M. Brewer\altaffilmark{2},
Stuart Lynn\altaffilmark{9},
Robert Simpson\altaffilmark{8},
Abe J. Hoekstra\altaffilmark{10},
Thomas Lee Jacobs\altaffilmark{10},
Daryll LaCourse\altaffilmark{10},
Hans Martin Schwengeler\altaffilmark{10},
Mike Chopin\altaffilmark{10}
Rafal Herszkowicz\altaffilmark{10} 
} 
\email{ji.wang@yale.edu}

\altaffiltext{1}{This publication has been made possible by the participation of 
more than 200,000 volunteers in the Planet Hunters project. Their contributions are 
individually acknowledged at http://www.planethunters.org/authors}
\altaffiltext{2}{Department of Astronomy, Yale University, New Haven, CT 06511 USA}
\altaffiltext{3}{NASA Ames Research Center, M/S 244-30, Moffett Field, CA 94035, USA}
\altaffiltext{4}{Bay Area Environmental Research Institute, Inc., 560 Third Street West, Sonoma, CA 95476, USA}
\altaffiltext{5}{Department of Physics, University of Notre Dame, 225 Nieuwland Science Hall, Notre Dame, IN 46556, USA}
\altaffiltext{6}{Department of Physics, Yale University, P.O. Box 208121, New Haven, CT 06520, USA}
\altaffiltext{7}{Yale Center for Astronomy and Astrophysics, Yale University, P.O. Box 208121, New Haven, CT 06520, USA}
\altaffiltext{8}{Oxford Astrophysics, Denys Wilkinson Building, Keble Road, Oxford OX1 3RH}
\altaffiltext{9}{Adler Planetarium, 1300 S. Lake Shore Drive, Chicago, IL 60605, USA}
\altaffiltext{10}{Planet Hunters}
\altaffiltext{11}{Institute for Astronomy, Department of Physics, ETH Zurich, Wolfgang-Pauli-Strasse 16, CH-8093 Zurich, Switzerland}

\begin{abstract}

We report the latest Planet Hunter results, including PH2~b,
a Jupiter-size (R$_{\rm{PL}}=10.12\pm0.56$ R$_\oplus$) planet orbiting 
in the habitable zone of a solar-type star. 
PH2~b was elevated from candidate status when a series of false positive tests 
yielded a 99.9\% confidence level that transit events detected around the 
star KIC~12735740 had a planetary origin. 
Planet Hunter volunteers have also discovered 42 new planet candidates 
in the $Kepler$ public archive data, of which 33 have at least three transits recorded. Most of these transit candidates have orbital periods longer than 100 days and 20 are potentially located in the habitable zones of their host stars. Nine candidates were detected with only two transit events and the prospective periods are longer than 400 days. The photometric models suggest that these 
objects have radii that range between Neptune to Jupiter. These detections 
nearly double the number of gas giant planet candidates orbiting at habitable 
zone distances. We conducted spectroscopic observations for nine of the brighter targets 
to improve the stellar parameters and we obtained adaptive optics imaging for 
four of the stars to search for blended background or foreground stars that 
could confuse our photometric modeling. We present an iterative analysis method 
to derive the stellar and planet properties and uncertainties by combining the available 
spectroscopic parameters, stellar evolution models, and transiting light curve 
parameters, weighted by the measurement errors. 
Planet Hunters is a citizen science project that 
crowd-sources the assessment of NASA $Kepler$ light curves. The discovery 
of these 43 planet candidates demonstrates the success of citizen
scientists at identifying planet candidates, even in longer period orbits 
with only two or three transit events.

\end{abstract}


\section{Introduction}
The past two decades have ushered in a new subfield in astronomy with the discovery of
exoplanets. More than 800 exoplanets have been detected as of October
2012~\citep{Wright2011, Schneider2011}. The majority of known exoplanets have been
discovered by one of two methods, the Doppler (radial velocity) technique or photometric
transit measurements. The combination of these two methods is particularly powerful since
it yields both planet mass and radius, which cannot be obtained by using either of the
methods alone. This information helps to reveal the planet composition 
and structure \citep{Madhu2012,Fortney2010}, and provides insight about planet formation 
and evolution \citep{Carter2012,Boue2012,Rice2012}.

The NASA $Kepler$ spacecraft was launched on March 7, 2009. The mission is monitoring
$\sim$160,000 stars with a relative photometric precision of $\sim$20 ppm in 6.5 h for
$K_p$=12 mag stars to search for exoplanets~\citep{Koch2010}. More than 2300 planet
candidates have been announced by analyzing the first 16 month $Kepler$
data~\citep{Batalha2012}. By most estimates, the fraction of actual planets among the 
planet candidates is greater than 90\% \citep{Lissauer2012,Morton2011}. Most of 
the $Kepler$ planet candidates have 
radii that are comparable to or smaller than Neptune~\citep{Borucki2011,Batalha2012}. 
If the planet masses are also comparable to Neptune, the induced stellar velocity 
amplitudes will be small, presenting an observational challenge for Doppler follow-up to measure 
the planet mass. As a result,
only $\sim$1 - 2\% of $Kepler$ planet candidates have been confirmed by radial velocity 
measurements. In some rare cases, transit timing variations (TTVs) of multi-planet systems 
have been modeled to confirm the planetary status of $Kepler$ 
candidates \citep{Holman2010,Lissauer2011,Ford2012}. Recently, 
statistical analysis has been used to estimate false positive probabilities for whether
a transiting object is planetary in nature; for very high confidence levels, candidates 
have been promoted to \textit{bona fide} planet status \citep{Fressin2013}.

The $Kepler$ team uses a wavelet-based algorithm
called transit planet search (TPS) \citep{Jenkins2002,Jenkins2010,Tenenbaum2012} 
to detect transits of exoplanets. Transit-like signals with a detection significance
greater than 7.1$\sigma$ and at least three transits are visually examined by the $Kepler$ team 
in order to confirm or reject spurious results. While the algorithms are 
effective and efficient, a small fraction of planet candidates can slip
past the TPS algorithm. In principle, human observers could also detect many of the transit
events; however, the sheer number of light curves is daunting. There are currently more 
than two million quarter-year light curves for the 160,000 stars in the $Kepler$ field. 

The Planet Hunters project\footnote{http://www.planethunters.org/} is one of the 
Zooniverse projects\footnote{https://www.zooniverse.org/} 
\citep{Lintott2008,Lintott2011,Fortson2012} and was launched on 2010 December 16 to 
search for transiting planets in the first quarter of $Kepler$ public archive 
data \citep{Fischer2012}. 
As of 2012 November, 13 quarters of data have been released by
the $Kepler$ Mission into a public archive hosted by the Mikulski Archive for Space Telescopes
(MAST\footnote{http://archive.stsci.edu}) at the Space Telescope Science Institute and by
the NASA Exoplanet Archive\footnote{http://exoplanetarchive.ipac.caltech.edu}.
These data are being processed on the Planet Hunters site and 
the collective of more than 200,000 volunteers around the globe has 
contributed $\sim$100 man-years to the visual inspection of $Kepler$ light 
curves. 

In this paper, we present an update of the research progress by the Planet Hunters participants: 
43 new planet candidates have been discovered, including 40 candidates with periods 
longer than 100 days. 
One of these objects has a planet confidence level of 99.92\%, high enough to be considered
a confirmed planet and we identify this object as PH2~b. At least three transits have 
been detected for 34 of the 
candidates. For nine planet candidates, only two transits were found in the light curves 
and a watch list has been established with the predicted time for expected future transit 
events.  Twenty of the candidates are potentially in the habitable zone and model fits 
of the light curves suggest that their radii range from a few to more than ten times 
the radius of the Earth (a Neptune to Jupiter range of radii).
We obtained follow-up observations for nine of the brighter stars including spectra from 
Keck HIRES~\citep{Vogt1994} to improve the stellar parameters and we obtained 
high spatial resolution 
imaging of four host stars using Keck NIRC2 adaptive optics~\citep{Wizinowich2000b}. 

We briefly introduce the Planet Hunters project in \S \ref{sec:PH} and describe our false positive tests in \S \ref{sec:FalsePosTest}, follow-up 
spectroscopy in \S \ref{sec:SpectrAnal} and orbit modeling in \S \ref{sec:iterative}.
In \S \ref{sec:NewCand}, we discuss the discovery of 43 new planet candidates, 
including a Jupiter-size planet orbiting in the habitable zone of its host star and 
new multi-planet systems. We discuss the overlap between these 
candidates and the transit crossing events (TCEs) identified by \citet{Tenenbaum2012}
in \S \ref{sec:Summary}. 

\section{Planet Hunters: A Citizen Science Partnership}
\label{sec:PH}

In the past two years, Planet Hunters have identified hundreds of planet
candidates that were also found by the $Kepler$ TPS algorithms. Because those 
detections had already been announced by the $Kepler$ team, we did not publish 
independent detections. However, a few unique systems were identified 
that had been overlooked by the $Kepler$ automatic detection and validation
pipeline.~\citet{Fischer2012} reported two planet candidates, KIC 10905746 b and KIC
6185331 b. ~\citet{Lintott2012} announced KIC 4552729 b, which 
shows evidence of transit timing variations, and
KIC~10005758~c, which is a second candidate to the host star of a $Kepler$-detected planet
candidate, KIC~10005758~b \citep{Batalha2012}. ~\citet{Schwamb2012} discovered PH1~b, a confirmed circumbinary
planet in a $\sim$137 day orbit around an eclipsing binary (EB) in a quadruple stellar
system. ~\citet{Schwamb2012b} carried out a systematic analysis and showed
that participants on the Planet Hunter site are most effective at detecting
transiting planet candidates with radii larger than 4 Earth radii; smaller planets 
tend to be lost in the noise and are best retrieved with mathematical algorithms. 

For each star in a $Kepler$ observation quarter, the light curve is cut into three equal pieces with 5 day overlap at the ends. This $\sim$30-day light curves are presented to participants for identifying possible planet candidate. Currently, data from Q1, Q2, Q3, Q4, Q5, and Q7 have been uploaded and screened. The inspections are randomly distributed among all stars with approximately 10 people viewing each light curve in Q1 and 5 people viewing each light curve in Q2, Q3, Q4, Q5 and Q7. Transit identifications by Planet Hunter volunteers are extracted from the 
database by weighting the results from multiple participants~\citep{Schwamb2012b}
and are then inspected by the science team. We used Q1, Q2, Q3, Q4, Q5 and Q7 to discover the candidates. After identification of planet candidates, we used Q1-Q13 data to derive orbital and stellar parameters for these systems. 

Another effective tool for identifying candidates is the object-orientated discussion, 
$Talk$\footnote{http://talk.planethunters.org - The code is available 
under an open-source license at https://github.com/zooniverse/Talk}, which is integrated 
into the Planet Hunters classification interface. $Talk$ is designed to enable 
discussion amongst Planet Hunters volunteers and to facilitate volunteer-led analysis 
of interesting light curves that can then be easily brought to the attention of 
the science team. Forum-style message boards are hosted on $Talk$ where
each $\sim$30-day light curve presented on the main Planet Hunters classification 
interface has a dedicated page. Volunteers can write comments, add searchable 
Twitter-like hash tags, create longer side discussions, and group similar light 
curves together in collections. Inside of $Talk$, 
Planet Hunters volunteers can examine all of the additional light curve 
data (not just the 30-day segment) available for interesting targets.

For Planet Hunters who opt to use $Talk$ after classifying a light curve, there is 
significant expertise available from a core of dedicated volunteers. Co-authors 
Kian Jek, Abe J. Hoekstra, 
Thomas Lee Jacobs, Daryll LaCourse, and Hans Martin Schwengeler have led an effort 
to collect transit candidates and to carry out transit modeling and initial 
vetting for false positives.
Their vetted lists contained the 43 candidates presented in this paper.

\section{Vetting the Transit Candidates: False Positive Tests}
\label{sec:FalsePosTest}

After transit candidates are detected by volunteers, the science 
team carries out an extensive effort to eliminate 
false positives that can occur from background blends or 
on-source grazing and eclipsing binaries 
\citep{Batalha2010, Prsa2011, Rowe2010}. A number of techniques 
have been developed to identify these interlopers \citep{Torres2011}. As a first step,
we reject light curves that have a V-shape, suggesting an eclipsing binary star. 
We examined all of the prospective transit curves for changing transit depths that can 
occur when a nearby star contributes a variable amount of 
flux to eclipsing binaries. Often the background variability is 
quarter-dependent, changing as the instrumental point 
spread function (PSF) varies. Another clear sign of an eclipsing binary 
star is variation in the transit depth for even and 
odd transits or a strong signal during secondary eclipse.  In addition to 
these checks, we developed code to check for pixel centroid offsets during 
the time of the prospective transit 
that are diagnostic of an off-target source. 
Although we were limited in our follow-up efforts by the 
availability of telescope time (for both spectroscopic and adaptive optics 
observations), we obtained follow-up observations for nine
of our transit candidates.

\subsection{Pixel Centroid Offset Analysis}

The $Kepler$ light curve files
contain information on both the position of the star at a given time (flux-weighted
centroids, called MOM\_CENTR1 and MOM\_CENTR2) and the predicted position the star should
be in based on the position of reference stars (called POS\_CORR1 and POS\_CORR2). By
subtracting one from the other we can find how much the position of the star differs from
its predicted position. We measured the flux difference during the prospective 
transit and immediately after transit to search for centroid shifts. If the transit 
occurs on a star other than the target star the centroid position
should move in the direction away from that star by an amount proportional to the transit
depth~\citep{Bryson2013}. The magnitude of the shift can be used to calculate
how far away this source is from the target. Table \ref{tab:orbital_params} reports the pixel centroid offset significance: the ratio of pixel centroid offset and the measurement error. Even if no obvious shift is seen, we can
still calculate a confusion region around the star. In our false positive analysis we use
the 3-$\sigma$ confusion radius as the outer limit on the separation between the target
and a false positive source. 

Pixelated flux time series measurements are available from the target pixel files. This
provides another false positive test. If flux changes take place outside $Kepler$
aperture, this is a strong indication that the observed flux change within aperture is a
result of contamination rather than planet transit. We used the PyKE analysis
tool~\citep{Barclay2012} for this analysis.

\subsection{Transit Depth Variation and Secondary Eclipse Search}

The light curve of an eclipsing binary will show different transit depths if the two stars
do not have identical radii. Thus, eclipsing binary stars are likely to exhibit
transit variations depending on whether the transiting body is the primary or secondary star. We
therefore search for possible transit depth variations between odd and even numbered
transits. No significant variations were detected in light curves for the planet candidate 
presented here.

For many eclipsing binaries, a secondary eclipse could also be detected since the 
stellar signal is so strong (compared to secondary eclipse of a planet). We perform a secondary
eclipse search using a phase-folded light curve with stellar photometric variability removed. PyKE~\citep{Barclay2012} is used to flatten and fold the light curve in phase. We only find one larger than 3-$\sigma$ detection, which is for KIC~4760478 at the phase of 0.437 (assuming primary transit takes place at the phase of 0) with a significance of 3.3 $\sigma$. The depth of the secondary transit for KIC~4760478 is 2518$\pm$758 ppm. Other than that, we do not detect any significant sign of a secondary eclipse signal, which is a necessary condition for the planetary origin of the
transit.  

\subsection{Adaptive Optics Imaging} 
To help rule out false-positive scenarios, we acquired high spatial resolution images 
for four stars: KIC~4820550, KIC~4947556, KIC~9958387, and KIC~12735740 using NIRC2 
(instrument PI: Keith Matthews) and the Keck II Adaptive Optics (AO) system \citep{Wizinowich2000}. 
The AO images were obtained on UT 2012
October 21 with excellent seeing of less than 0.3 arcsec. Our observations consist of dithered frames obtained with the 
KÕ filter. We used the narrow camera setting (10 mas $\mbox{pix}^{-1}$ plate scale) to 
provide fine spatial sampling of the instrument point-spread function. For each star, 
we acquired at least 90 seconds of on-source integration time. Complementary data 
taken with the J-filter helped to distinguish between speckles and off-axis sources 
in several cases. 

\begin{figure}
\begin{center}
\includegraphics[width=16cm,height=16cm]{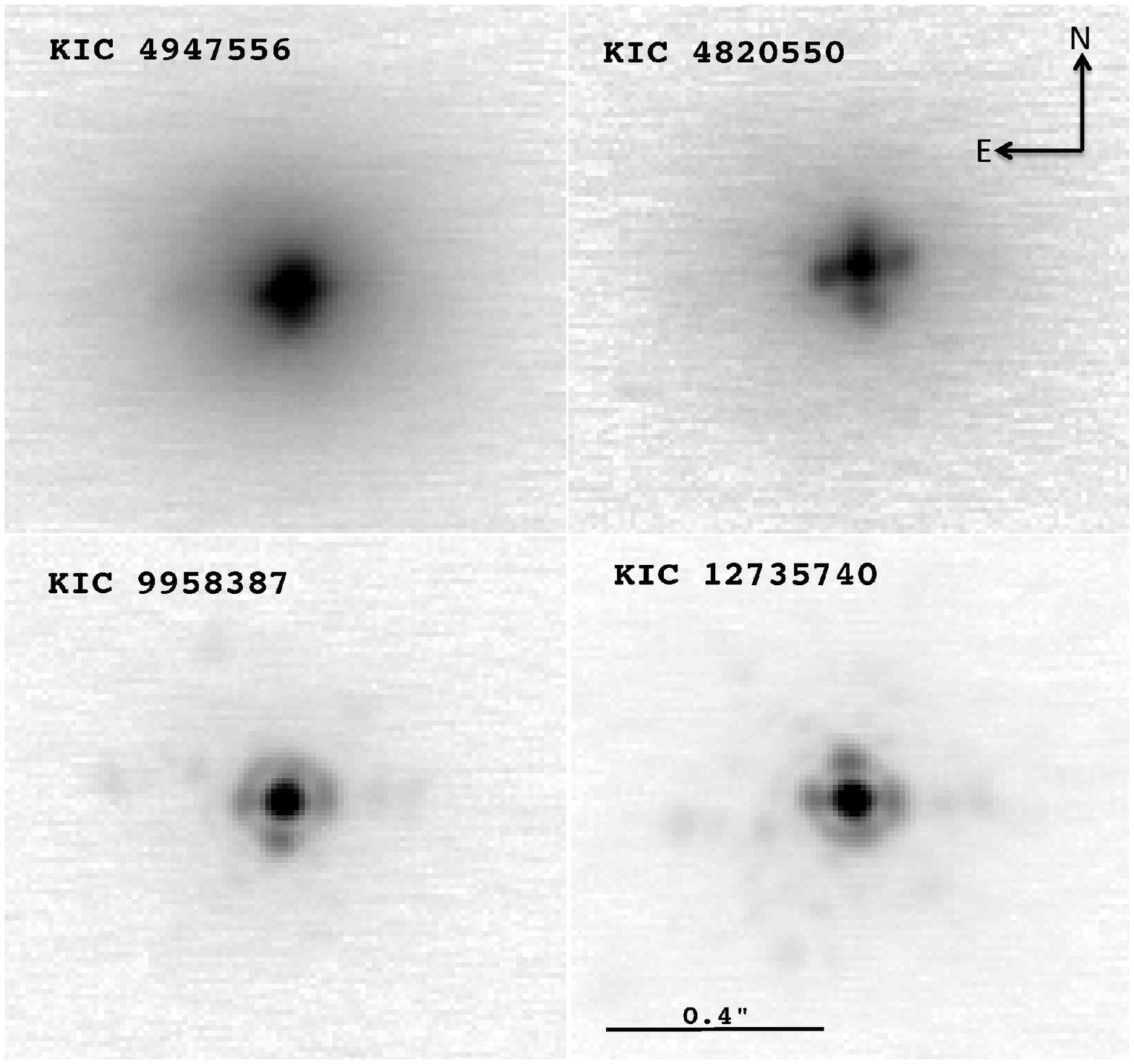} \caption{AO images for KIC~4947556, KIC~4820550, KIC~9958387 and KIC~12735740. North is up
and east to the left. Horizontal bar is about 0.4 arcsec. Negative square root scale is
used to show off-axis features. No additional sources were observed down to the magnitude
limits listed in Table \ref{tab:AO_params}. \label{fig:Keck_AO}}
\end{center}
\end{figure}

Raw NIRC2 data was processed using standard techniques to replace hot pixels, 
flat-field the array, subtract thermal background radiation, and align and co-add 
frames. The primary star was not saturated in any individual image. We did not find 
any evidence for neighboring sources in either the raw or processed data 
(Figure \ref{fig:Keck_AO}). Our diffraction-limited 
observations rule out the presence of companions and background sources as listed in 
Table \ref{tab:AO_params} for several characteristic angular separations. Contrast is calculated by 
measuring the flux from scattered light relative to the stellar peak. Specifically, 
we calculate the standard deviation of intensity inside a box of size 3 x 3 FWHM 
(full-width at half-maximum), where the FWHM corresponds to the spatial scale of 
a speckle ($\approx 45$ mas). The results from each box are then azimuthally averaged 
to produce a radial profile. We nominally achieve (at least) 4 magnitudes of 
contrast ($10\sigma$) at 0.5$^{\prime\prime}$.

\section{Stellar Properties}
\label{sec:SpectrAnal}

For most of the planet candidates, we adopted stellar parameters from the $Kepler$ 
Input Catalog (KIC). According to~\citet{Brown2011}, stellar radii from KIC have
typical error bars of 0.2 dex, corresponding to a percentage error of 58\%; typical
error bars are 0.5 dex for log $g$ and 200 K for T$_{\rm{eff}}$. 

For one star, KIC~8012732, there were no available KIC parameters so
we estimated the effective temperature for this star from the $g-r$ color, using the 
color-temperature relations from~\citet{Pinsonneault2012}. We do not know the 
amount of reddening for this star and we found that adopting reddening values in
the range 0.0 to 0.1 translates 
to a large uncertainty for $T_{\rm eff}$ values, from 5536 K to 6517 K. We adopted 
an intermediate $T_{\rm eff}$ of 6000~K with an error bar of 500 K. We also arbitrarily 
adopt a corresponding main sequence value for log $g$ of $4.3\pm0.5$ and 
[Fe/H] of $0.0\pm0.5$.

\subsection{Spectroscopic analysis}
\begin{figure}
\begin{center}
\includegraphics[width=16cm,height=12cm]{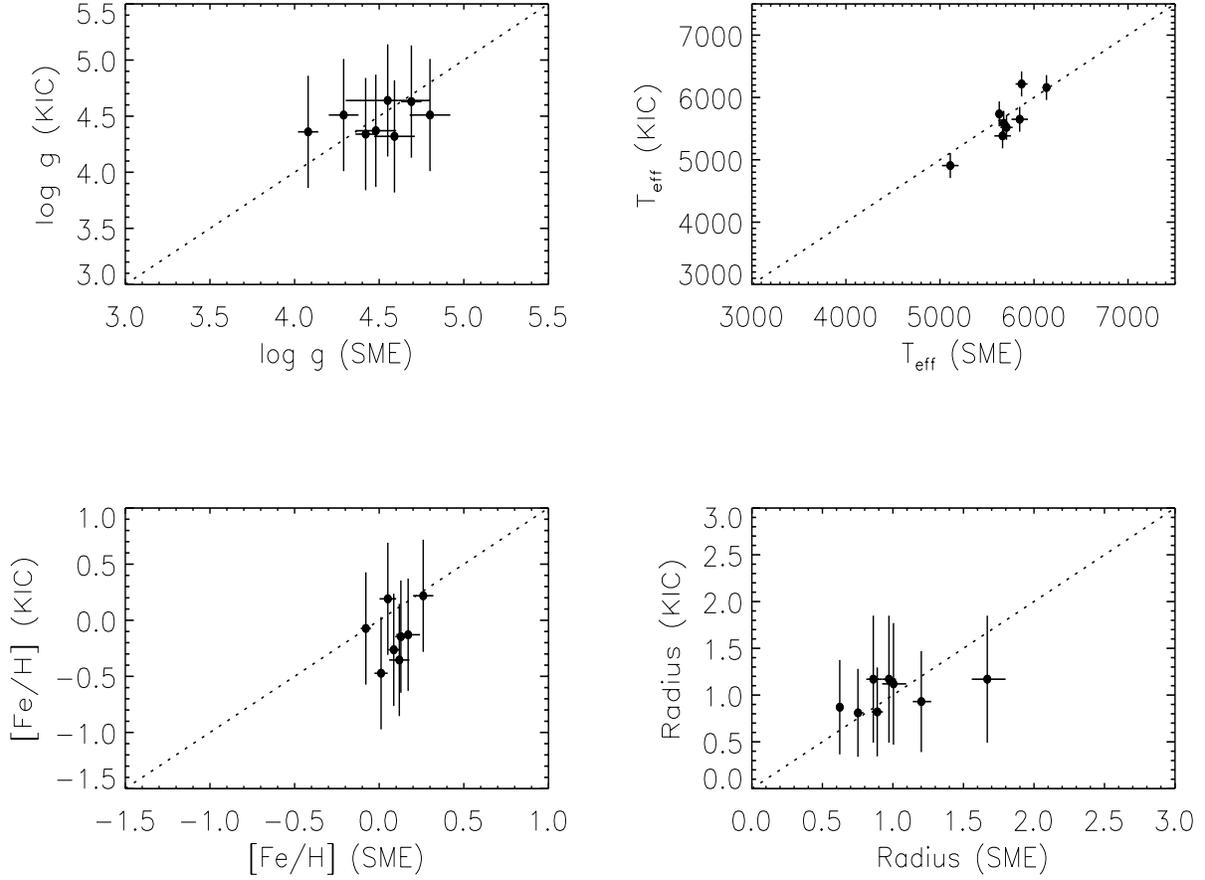} 
\caption{Comparison of log $g$, $T_{\rm{eff}}$, [Fe/H] and stellar radius between 
SME and KIC values for 9 stars with Keck HIRES spectra. The parameters are consistent with each other 
within reported error bars. Stellar radii are inferred based on Yonsei-Yale isochrone 
interpreter~\citep{Demarque2004}. 
\label{fig:comp_SME_KIC}}
\end{center}
\end{figure}

We obtained
spectra for nine relatively bright stars using Keck HIRES \citep{Vogt1994} on 
UT 2012 October 07: KIC~2581316, KIC~4472818, KIC~4820550, KIC~4947556,
KIC~9147029, KIC~9958387, KIC~11253827, KIC~11392618 and KIC~12735740. The spectra have a
resolution of $\sim$55,000 with signal-to-noise ratios typically greater than $\sim$50 at
around 5160 $\AA$. We used Spectroscopy Made Easy (SME) \citep{Valenti1996,Valenti2005} 
to model our observed spectra and to derive 
best-fit stellar parameters for $T_{\rm{eff}}$, log $g$, [Fe/H], [$\alpha$/Fe] and
$v\sin i$. The parameter uncertainties were estimated by running a grid of models 
with different initial guesses for $T_{\rm{eff}}$ and either using the standard deviation 
of the output model parameters or the error analysis of \citet{Valenti2005} (whichever was larger). 
The SME results are summarized in Table \ref{tab:SME_params}.
Figure \ref{fig:comp_SME_KIC} shows the comparison of log $g$, $T_{\rm{eff}}$ and 
[Fe/H] between SME results and KIC values. They are generally in agreement within 
reported error bars. 

Although we have metallicity measurement precisions ranging from 0.03 to 0.09 dex 
for nine stars with SME analysis, given the median 
measurement uncertainty of $\sim$0.5 dex from the KIC for other host stars, no 
definitive conclusion can be made with regards to correlation between planet 
occurrence and stellar metallicity.

\subsection{Isochrone interpolation}
We used KIC values or (when available) the output of SME as input for Yonsei-Yale Isochrone
interpolation \citep{Demarque2004} in order to infer stellar properties such as mass,
radius and luminosity. A total of 1000 Monte Carlo trials were run using 
input parameters ($T_{\rm{eff}}$, log $g$, [Fe/H] and [$\alpha$/Fe]), randomly drawn 
from a Gaussian distribution scaled to the standard deviation of stellar parameter 
uncertainties. Stellar age and [$\alpha$/Fe] are also required inputs for the isochrone 
interpolation. We included a wide stellar age range of 0.08 to 15 Gyr and 
a range in [$\alpha$/Fe] of 0.0 to 0.5 for our Monte Carlo trials in order 
to explore a large parameter space. 
The lower right panel of Figure \ref{fig:comp_SME_KIC} shows the comparison between 
Yonsei-Yale model inferred stellar radii and those from KIC. The stellar radii from 
these two sources are generally consistent with each other within error bars.
Stellar mass, radius and luminosity were then inferred based on the 
mean and standard deviation of the outputs from the Yonsei-Yale Isochrone interpolation. 

\subsection{Chromospheric activity and photometric activity}
\begin{figure}
\begin{center}
\includegraphics[width=10cm,height=20cm]{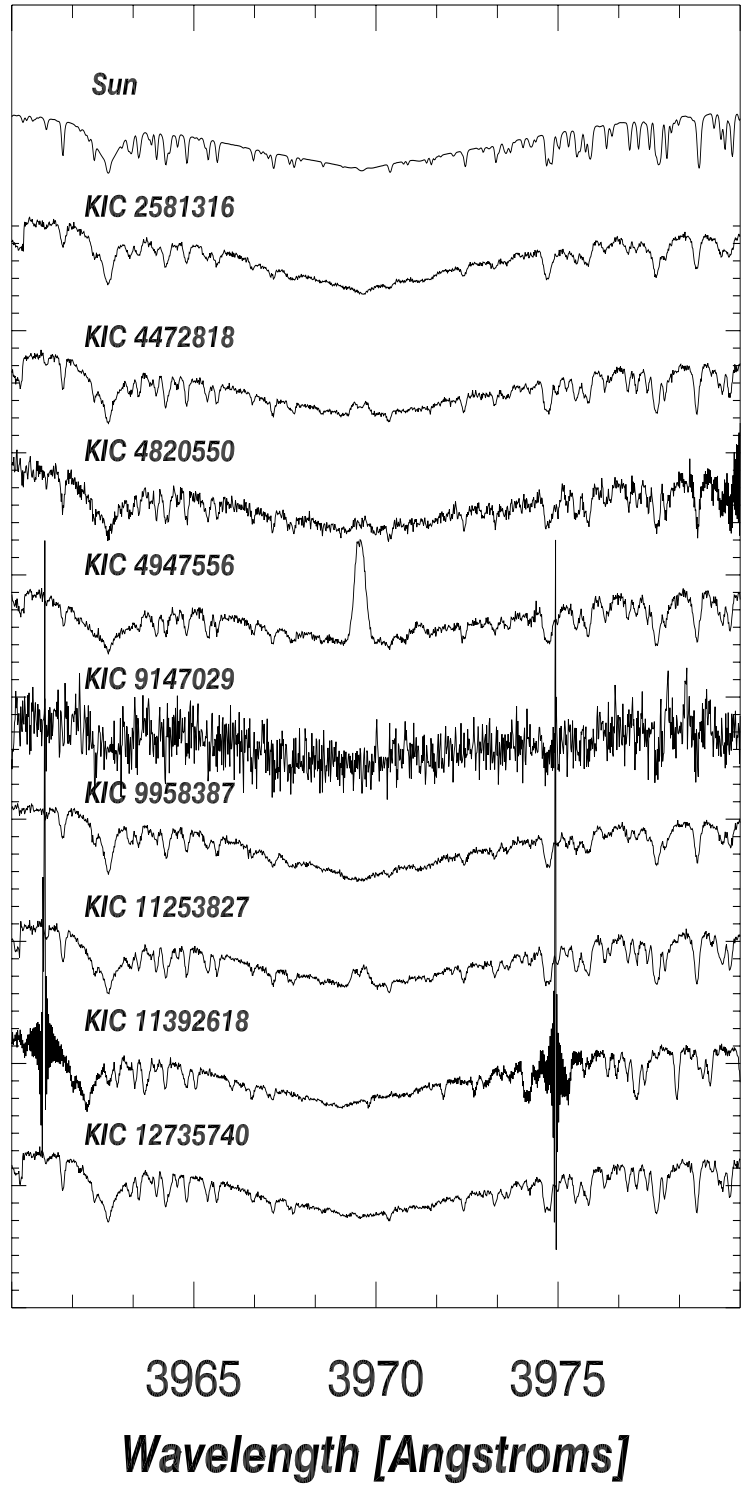} \caption{Ca
II H line (3968.5 $\AA$) for the Sun and our nine stars with Keck spectra: 
KIC~2581316, KIC~4472818, KIC~4820550, 
KIC~4947556, KIC~9147029, KIC~9958387, KIC~11253827, KIC 11392618, and KIC~12735740. 
The Ca II H\&K lines are a diagnostic for chromospheric activity
and are tied to the stellar age. \label{fig:CaHK}}
\end{center}
\end{figure}

The Ca II H\&K core emission is 
commonly used as a diagnostic of chromospheric activity in 
stars~\citep{Gray1985,Gunn1998,Gizis2002,Isaacson2010} and this feature 
is included in the spectral format of our Keck HIRES observations. To account for 
different continuum flux levels near the Ca II lines for stars of different spectral 
types, the core emission is often parameterized as \rhk, a logarithmic fraction 
of the flux in the H \& K line cores to photospheric contributions from the star. 
Chromospheric activity declines with stellar age (typically, subgiants show very 
low activity), so cluster stars of known ages have been used to calibrate
\rhk\ to rotation periods and ages~\citep{Noyes1984}. Because the Kepler field stars 
tend to be faint and the obtained spectra have low signal-to-noise ratios, our usual pipeline analysis of chromospheric activity often fails. 
Therefore we used stars on the California Planet Search (CPS) programs as a 
comparison library. Matching with stars with similar effective temperatures, we carried out 
a visual inspection of the Ca II H\&K core emission to estimate \rhk\ for the 
set of nine stars with HIRES spectra. 
Figure \ref{fig:CaHK} shows our Keck HIRES spectra centered on the 
Ca II H lines (3968.5 $\AA$) for the Sun and the nine stars observed at Keck. 

In addition to Chromospheric activity, we checked the photometric activity of the stars in our sample. The light curve of each quarter was normalized to its median and then stitched using PyKE~\citep{Barclay2012}. The photometric variability is indicated by the ratio of out-of-transit flux standard deviation and the median flux. The adopted stellar parameters and properties for all 43 stars are summarized in Table \ref{tab:stellar_params}.

\section{An Iterative Approach to Stellar and Planet Property Determination}
\label{sec:iterative}
$Kepler$ light curves for quarters $1 - 13$ were de-trended and 
normalized for all planet candidates that passed the false positive tests. 
The light curves for these 43 stars were phase-folded and fitted with a model 
described by~\citet{Mandel2002}. 
The free parameters in our model include orbital period, eccentricity $e$, argument
of periastron $\omega$, inclination $i$, the ratio of semi-major axis and stellar radius
$a/\rm{R}_\ast$, the radius ratio of planet and star $\rm{R}_{\rm{PL}}/\rm{R}_\ast$, mid
transit time, linear limb darkening and quadratic limb darkening parameters. A
Levenberg-Marquardt least squares algorithm was used to find the best-fit parameters. 
Uncertainties associated with parameters were estimated with a
bootstrapping process in which the data points in transit light curves are perturbed based
on photometric measurement uncertainty and the initial guess is perturbed based on
standard deviation of previous runs.

\begin{figure}
\begin{center}
\includegraphics[angle=0, width=1.0\textwidth]{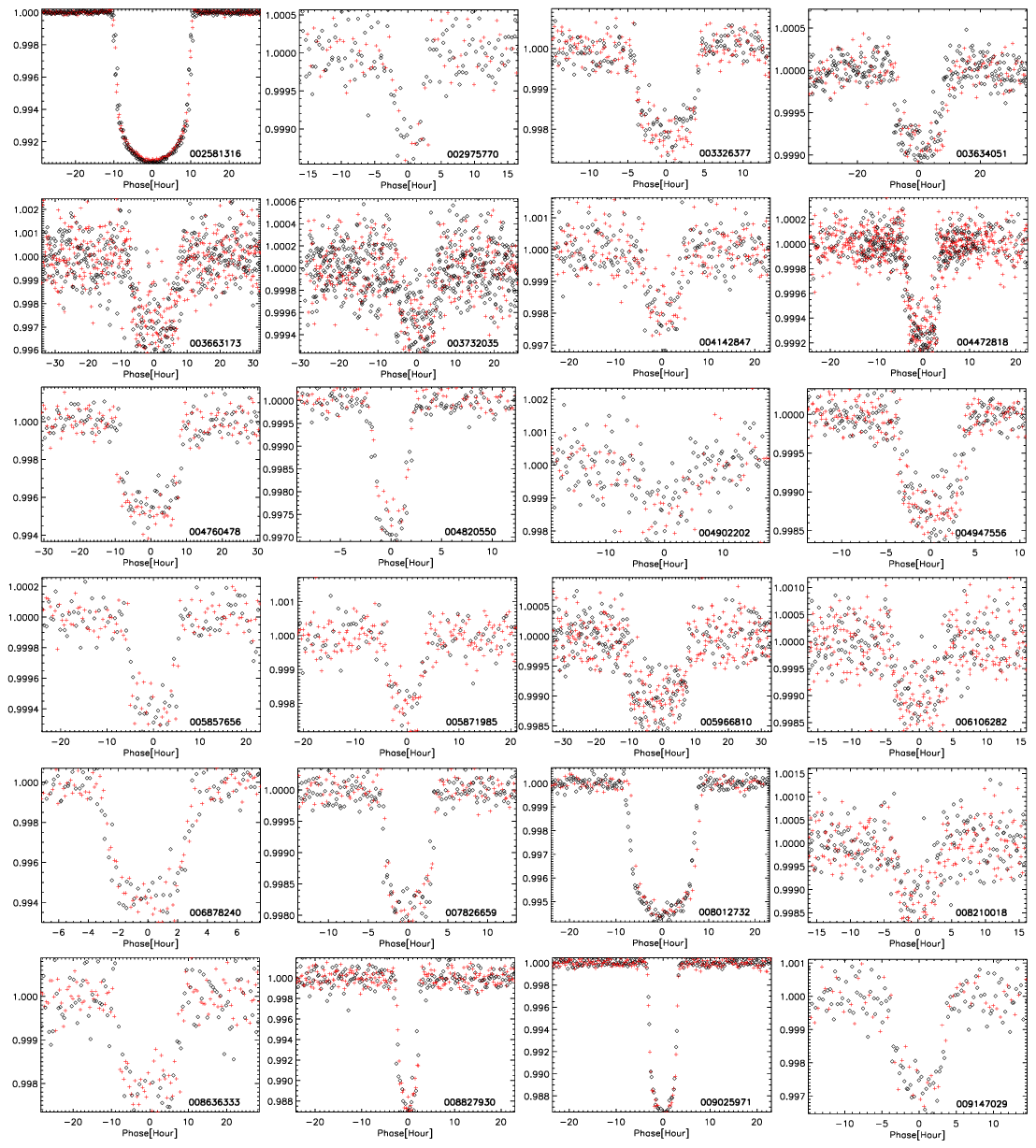} 
\caption{Phase folded light curves for planet candidates discovered by the Planet
Hunters project. Stellar and orbital information can be found in Table
\ref{tab:stellar_params} and \ref{tab:orbital_params}. Black diamonds are data points from
odd number transits and red pluses are data points from even number transits.  
\label{fig:PH_HZ_1_2}}
\end{center}
\end{figure}

\begin{figure}
\begin{center}
\includegraphics[angle=0, width=1.0\textwidth]{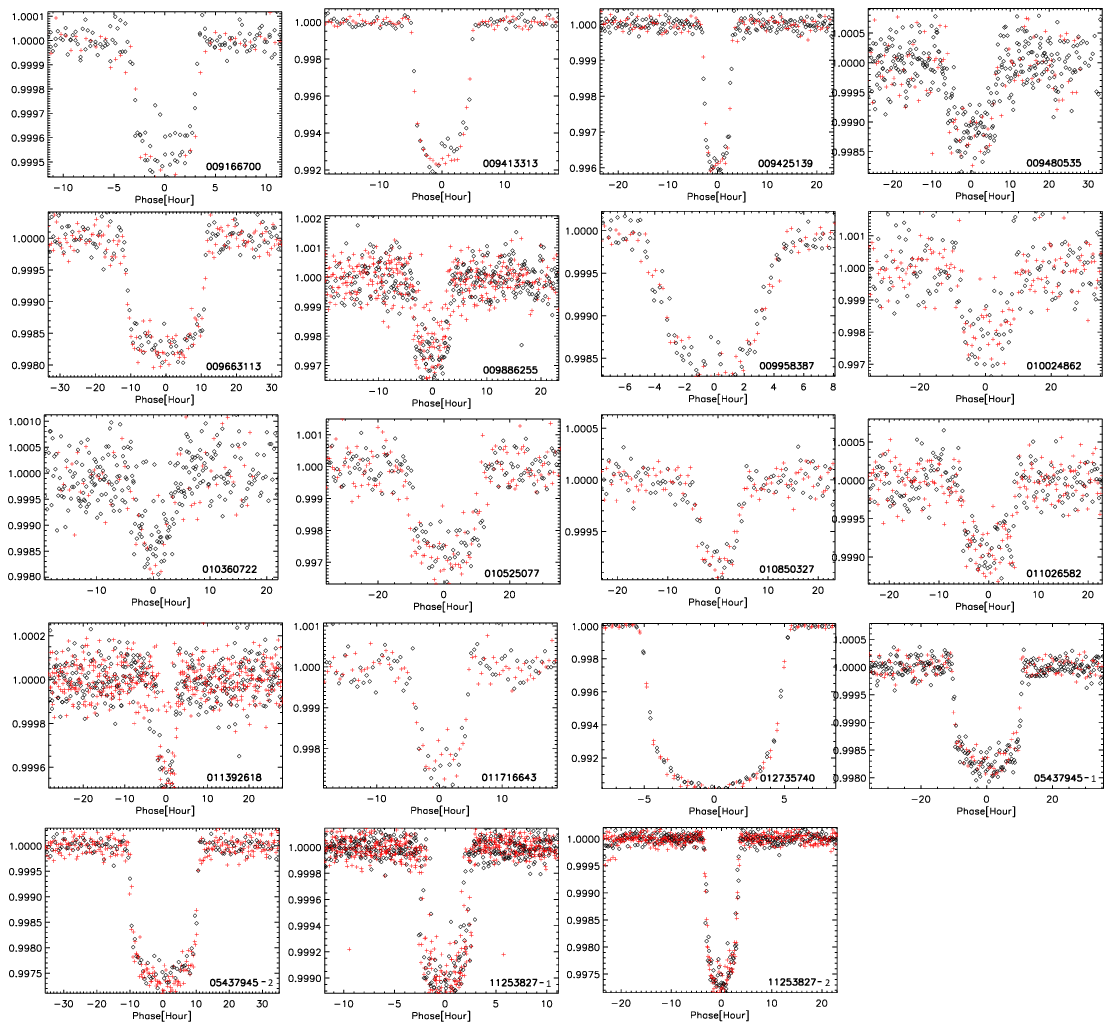} 
\caption{Phase folded light curves for planet candidates discovered by the Planet
Hunters project. Stellar and orbital information can be found in Table
\ref{tab:stellar_params} and \ref{tab:orbital_params}. Black diamonds are data points from
odd number transits and red pluses are data points from even number transits. 
\label{fig:PH_HZ_1_2_4}}
\end{center}
\end{figure}

There are often several types of data that offer constraints on the stellar
attributes and orbital parameters.  Unfortunately, the data quality is uneven 
and some data may be missing altogether. 
We developed a code to consider the data quality and to combine all available information
in order to obtain self-consistent determinations of stellar and orbital parameters 
along with uncertainties. We iterate between a photometric determination of 
$a/\rm{R}_\ast$ and an 
estimate of $a/\rm{R}_\ast$ derived from the stellar parameters. Time-series photometry observations yield a distribution for the value of
$a/\rm{R}_\ast$ that we derive from the transit light curve model. The uncertainty of the stellar parameters
and the quality of the mass and radius from the isochrone analysis lead to 
a separate distribution for the ``evolution-model-determined"
value for $a/\rm{R}_\ast$. One set of $a/\rm{R}_\ast$ distribution helps to constrain parameter space that leads to the other set of $a/\rm{R}_\ast$ distribution. After few iterations, two sets of $a/\rm{R}_\ast$ distribution converge and a self-consistent stellar and orbital solution is thus obtained. This code was used to evaluate all planet candidates in
this paper and the derived orbital parameters are listed in Table \ref{tab:orbital_params}
and the phase-folded light curves are presented in Figures \ref{fig:PH_HZ_1_2} and 
\ref{fig:PH_HZ_1_2_4}. 

\begin{figure}
\begin{center}
\includegraphics[width=16cm,height=12cm]{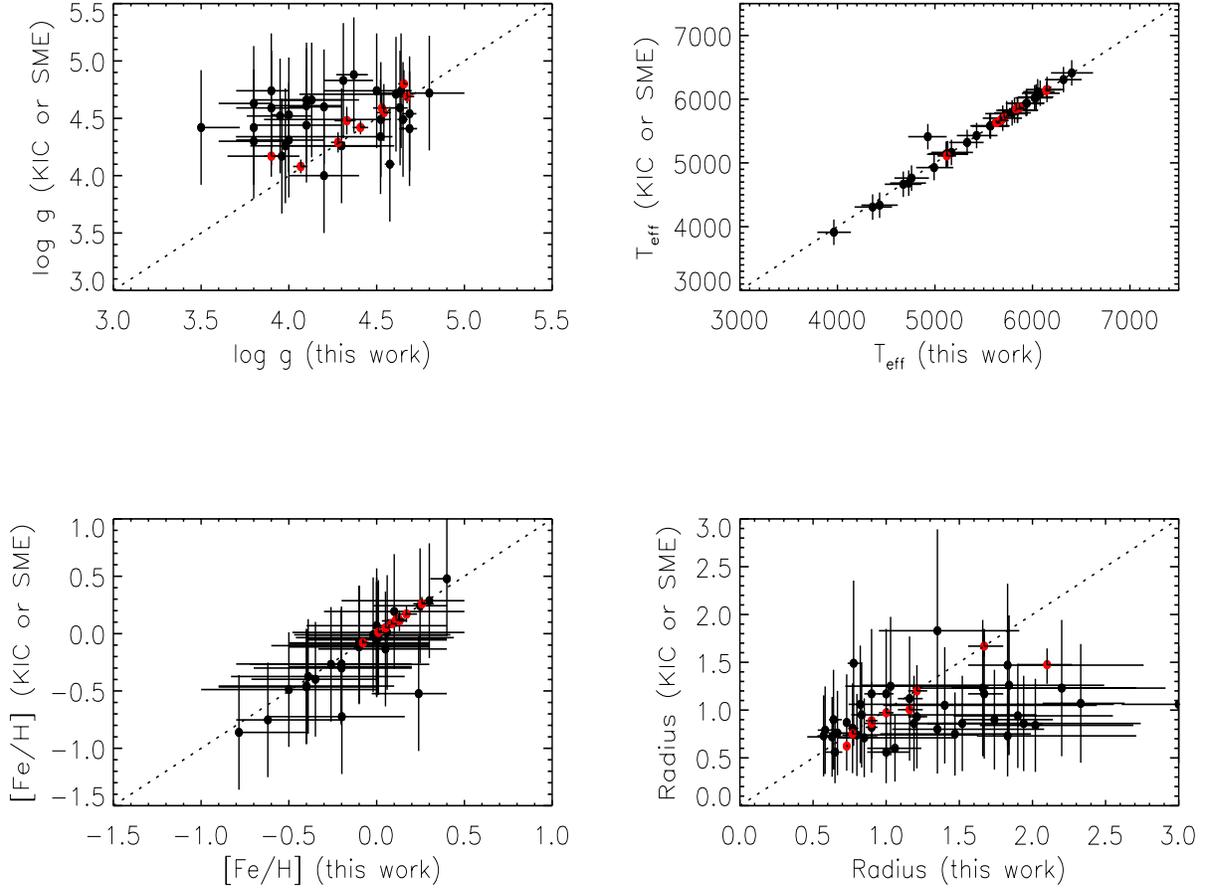} \caption{Comparison 
of log $g$, $T_{\rm{eff}}$, [Fe/H] and stellar radius between results from 
iterative algorithm (see \S \ref{sec:iterative} for details) and KIC (marked 
as black dots) or (when available) SME values (marked as red dots). The
log $g$ from KIC appear to be systematically overestimated and likely 
explains why stellar radii from KIC are smaller than the values 
we derive.  \label{fig:comp_iter_KICSME}}
\end{center}
\end{figure}

Figure \ref{fig:comp_iter_KICSME} shows the comparison of 
log $g$, $T_{\rm{eff}}$, [Fe/H] and stellar radius between the iterative 
method and KIC or SME values (when available). $T_{\rm{eff}}$ and [Fe/H] 
are generally consistent except for one outlier, KIC~4902202. KIC~4902202 
is at the fainter end of our planet candidates with a Kep mag of 15.58. 
We suspect that the stellar characterization error is probably due to the 
faintness of the star. The tendency for overestimated log $g$ values in 
the KIC catalog has been noted by~\citet{Brown2011} and~\citet{Verner2011},
this results in underestimated stellar radii in the KIC data. 

\subsection{Equilibrium Temperature}
The planet candidate equilibrium temperature is calculated after running the
iterative code discussed above. Two parameters, $T_{\rm{eff}}$ and $a/\rm{R}_\ast$, are
used to infer the equilibrium temperature with the following equation: \begin{equation}
\rm{T}_{\rm{PL}}=\rm{T}_{\rm{eff}}\cdot\left(\frac{1-\alpha}{4\varepsilon}\right)^{\frac{1}
{4}}\cdot\left(\frac{a}{R_\ast}\right)^{-\frac{1}{2}}, \end{equation} where $\alpha$ is
albedo and $\varepsilon$ is emissivity. We adopt $\alpha$ = 0.3 and $\varepsilon$ = 0.9,
which are typical values for Neptune and Jupiter in the solar system. Because orbital
eccentricity is barely constrained for most of the systems, zero eccentricity is assumed
in calculation; a moderate eccentricity does not significantly change the value of
$\rm{T}_{\rm{PL}}$ but changes the uncertainties.

\begin{figure}
\begin{center}
\includegraphics[width=16cm,height=10cm]{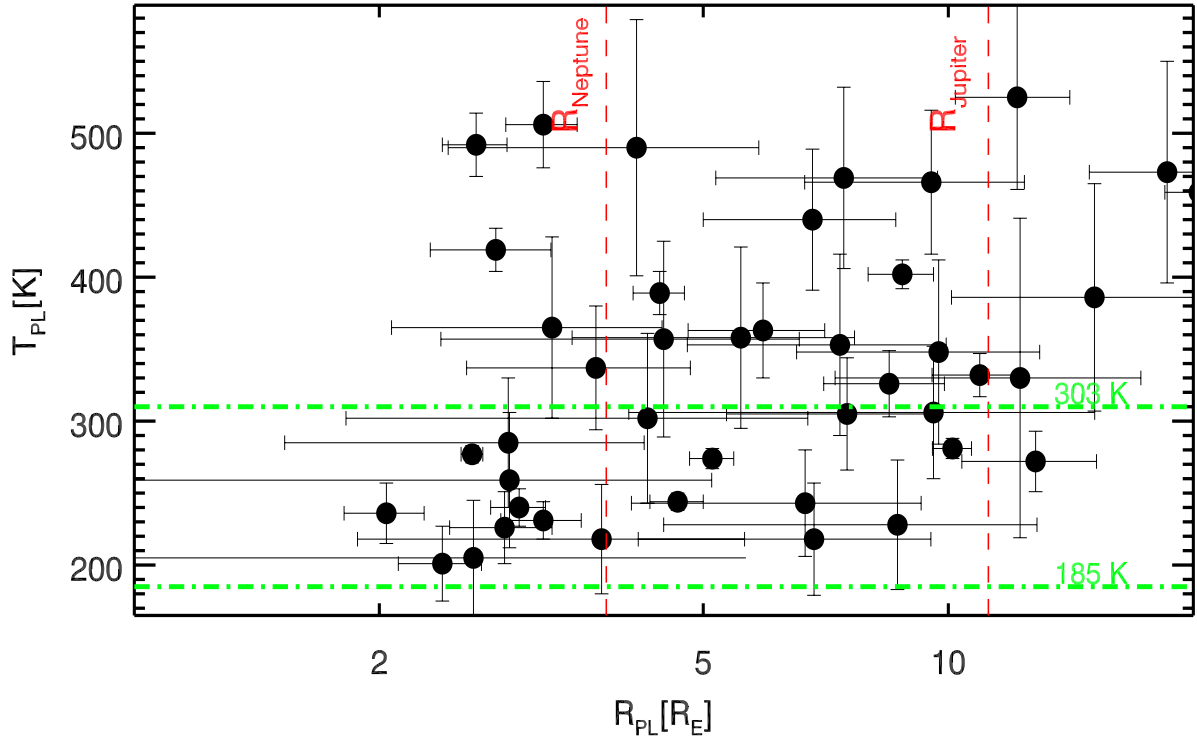} \caption{Equilibrium temperature vs. 
planet radius digram for the 43 planet candidates.
The boundaries for Neptune and Jupiter radii are plotted as red dashed lines and 
the hot and cold boundaries of habitable zone temperatures are limited by the green
dash-dotted lines. Most of the planet candidates are gas-giant planets and 20 
candidates appear to reside in the habitable zone 
(as defined in~\citet{Batalha2012}).  
\label{fig:Rp_Tpl}}
\end{center}
\end{figure}

Figure \ref{fig:Rp_Tpl} shows the distribution of planet candidates 
with equilibrium temperatures that range from 218~K to 525~K.  Depending on 
the orbital eccentricity or atmospheric albedo or emissivity, some of these 
planet candidates may have equilibrium temperatures consistent with habitable planets 
for at least a fraction of their orbits. The smallest of 
these candidates is KIC~4947556~b ($R_{P}=2.60\pm$0.08  R$_\oplus$, 
T$_{\rm{PL}}=277\pm6 K$). Eight other candidates are also 
smaller than Neptune but have larger uncertainties in their radii. The 
median metallicity (i.e., [Fe/H]) of the stellar hosts is -0.08 dex. 

\section{New Planet Candidates}
\label{sec:NewCand}

Table \ref{tab:orbital_params} contains only three planet candidates with 
orbital periods shorter than 100 days: KIC~4472818 (70.97~d) and two 
planet candidates around KIC~11253827 (44.99~d and 88.5~d).

Forty of our planet candidates have orbital periods longer than 100 days. 
We note that nine of these candidates have only two transits detected in the 13 quarters of 
$Kepler$ data and so would not meet the detection criteria of TPS 
algorithm \citep{Tenenbaum2012}, which used Q1-12. The 
derived orbital period for these nine candidates yields a prediction for when 
the next transit should occur 
and these objects are on a watch list on the Planet Hunters web site, however 
with only two transits, the orbital parameters for these candidates may have 
significant uncertainties.  

Long period candidates are particularly interesting because these planets may 
reside in or near the so-called habitable zone where liquid water could pool 
on the surface of a rocky planet. The temperature range defining the habitable zone 
is adopted from~\citet{Batalha2012} as T$_{\rm{PL}}$ between 185 and 303~K. 
Excluding the nine candidates with only two observed transits (and therefore with 
relatively large uncertainties in the prospective orbital parameters), 
we find that 15 of our candidates may reside in habitable zone orbits. 
Of course, these planets may not be habitable, especially since the modeled 
radii suggest that they are likely gas giant planets. 
However, these 15 detections add to the 19 planet candidates 
announced by~\citet{Batalha2012} with radii between Neptune and Jupiter and 
nearly double the set of gas-giants planet candidates 
in the habitable zone.

\subsection{Planet Confidence Levels}
Four of our stars have high quality follow-up (spectroscopic analysis and AO imaging) that 
allows us to quantify the false positive probabilities for planet candidates using a method 
called planetary synthesis validation (PSV). The PSV technique is based on the work 
of \citet{Fressin2012} and \citet{Barclay2013a} and has been used to confirm 
substellar planet candidates (Barclay et al. 2013b).
The goal is to restrict the parameter space where false
positives can exist and then to quantify the probability that we find a false positive in
this space. This false positive probability is then compared in a Bayesian manner to the
probability of finding a planet of the size measured, known as the planet prior. 
The ratio of the planet prior to the sum of false positive probability and the 
planet prior yields a confidence level for whether a particular source is a \textit{bone fide} planet.
A specific example is PH2~b, presented in \S \ref{sec:PH2}.

The parameter space we consider is a two dimensional plane with axes of projected spatial
distance between our target source and any false
positive source, which we refer to as separation, and the difference in 
brightness between our target and a false positive
source, which we refer to as $\Delta K_{P}$. We primarily consider background and
foreground systems that are either eclipsing binaries (EBs) or background transiting 
planet host stars. False positive scenarios involving a stellar hierarchical triple systems 
rarely provide a good fit to the observed light curve~\citep{Fressin2012} and 
are not considered here. As our goal here is
to determine the probability that the that the source we detect is substellar in nature
and physically associated with the host star, we also do not consider planets orbiting a
physical companion to be false positives in this study.

\subsection{Using Galactic Models to Quantify False Positive Rates} 
\label{sec:fap} 

We use the Bessan\c{c}on Galaxy model simulation~\citep{Robin2003} to estimate the number of
undetected background and foreground stars. First, we calculate a model population of
stars within 1 deg$^{2}$ of the target star. Then, for each $\Delta K_{P}$ not excluded by
the CCF or by transit depth, in steps of 0.5 in $K_{P}$, we count how many stars are in
the 1 deg$^{2}$ box. For example, for $\Delta K_{P}=3.0$, we count how many stars have a
$\Delta K_{P} = 3.0\pm0.25$. By using the separation constraints from centroids and AO
imaging we scale the number within 1 deg$^{2}$ to find the number within our confusion
limit using the equation \begin{equation} N_{\textrm{sep}} =  \pi  (S/3600)^{2}
N_{\textrm{deg}^{2}} \end{equation} where $S$ is the separation in arcseconds and the
number of stars within 1 deg$^{2}$ is $N_{\textrm{deg}^{2}}$. Then the total number of
background stars is found by integrating over the individual $\Delta K_{P}$ bins.

\subsection{Calculating the Planet Prior and Confidence in the Planetary Interpretation} 
In order to determine a confidence level that a particular transit
is planetary in nature, we must compare
this to the probability of finding a real planet. This planet
prior \citep{Fressin2012,Barclay2013a} makes use of the occurrence and false positive rates 
derived by \citet{Fressin2013} for various
sizes ranges of planets. For example, there are 223 giant planets ($6.0 <R_{\oplus}<22.0$)
in the $Kepler$ planet catalog and the false positive rate is 17.7\%. There is no
correction for completeness because in the case of large planets, all are expected to be
detected~\citep{Fressin2013}. Therefore, there the occurrence rate for giant planets is
$(1 - 0.177)*223 = 184$ giant planets per $1.6\times 10^{5}$ stars 
(the number of stars observed by $Kepler$).

The confidence level that a particular source is a \textit{bona fide} 
planet is determined by the ratio of the planet prior to the sum of false positive 
probability and the planet prior. We use the $Kepler$
eclipsing binary catalog~\citep{Slawson2011} and planet catalog~\citep{Batalha2012} and
assume all of the sources in these catalogs are either detached EBs on the target 
source (not background EBs) or transiting planets. We do not include contact binaries, which 
do not mimic planetary transit light curves. We find that 2.6\% of stars host are either 
EB's or hosts of transiting planet candidate. The planet confidence is calculated as:
\begin{equation} 
\label{eq:pc}
C = pp / ((fp_{stars} * 0.026 * N_{stars}) + pp) 
\end{equation}
where pp is the planet prior (e.g., 184 for giant planets), $fp_{stars}$ is the 
false positive rate, and the number of stars is $N_{stars} = 1.6 \times 10^5$. 

We were able to derive confidence levels for the four candidate planets where 
both AO imaging and spectroscopic analysis was carried out on the host stars: 
KIC~4820550 (88\%), KIC~4947556 (76\%) KIC~9958387 (86\%) and KIC~1235740 (99.92\%).
The high confidence rate for KIC~1235740 seems sufficient to suggest that the 
light curve dips are caused by a transiting exoplanet. These four systems are 
discussed in more detail subsequently. 

\subsection{PH2~b (KIC~12735740)}
\label{sec:PH2}

KIC~12735740 is relatively bright with a $Kepler$ magnitude of 12.62. 
The background photometric variability 
in the $Kepler$ light curves is less than 0.5\% and our evaluation of the emission 
in the cores of the Ca II H line for KIC~12735740 suggests that this is a 
chromospherically inactive star with an estimated \rhk\ $\le -4.9$.
Our Keck HIRES spectrum have a signal-to-noise ratio of $\sim$50 and we derive 
$T_{\rm{eff}}$ = $5629^{+42}_{-45}$~K and log $g$ = $4.408 \pm -0.044$ 
for KIC~12735740. The star has a roughly solar metallicity with 
[Fe/H] = -0.07 $\pm 0.05$ and is slowly rotating 
with v$\sin i$ = 1.43 $\pm$ 0.8 km $\rm{s}^{-1}$. 

Four transits of KIC~12735740 were detected by Planet Hunters with a period of 
282.5255 $\pm$ 0.0010~d, starting on the MJD of 55195.5709. Our best fit model 
for the transit light curves suggest that this is a Jupiter-size planet 
($10.12\pm0.56$ R$_\oplus$) with a temperature at the top of its atmosphere of  
T$_{\rm{PL}}$ = 281 $\pm$ 7~K. This gas giant planet appears to reside in the 
habitable zone as defined by the temperature limits of \citet{Batalha2012}.

We obtain a 99.92\% confidence level for a planetary nature and therefore interpret 
this as a \textit{bona fide} planet and assign it the designation PH2~b.

\begin{figure}
\includegraphics[width=\textwidth]{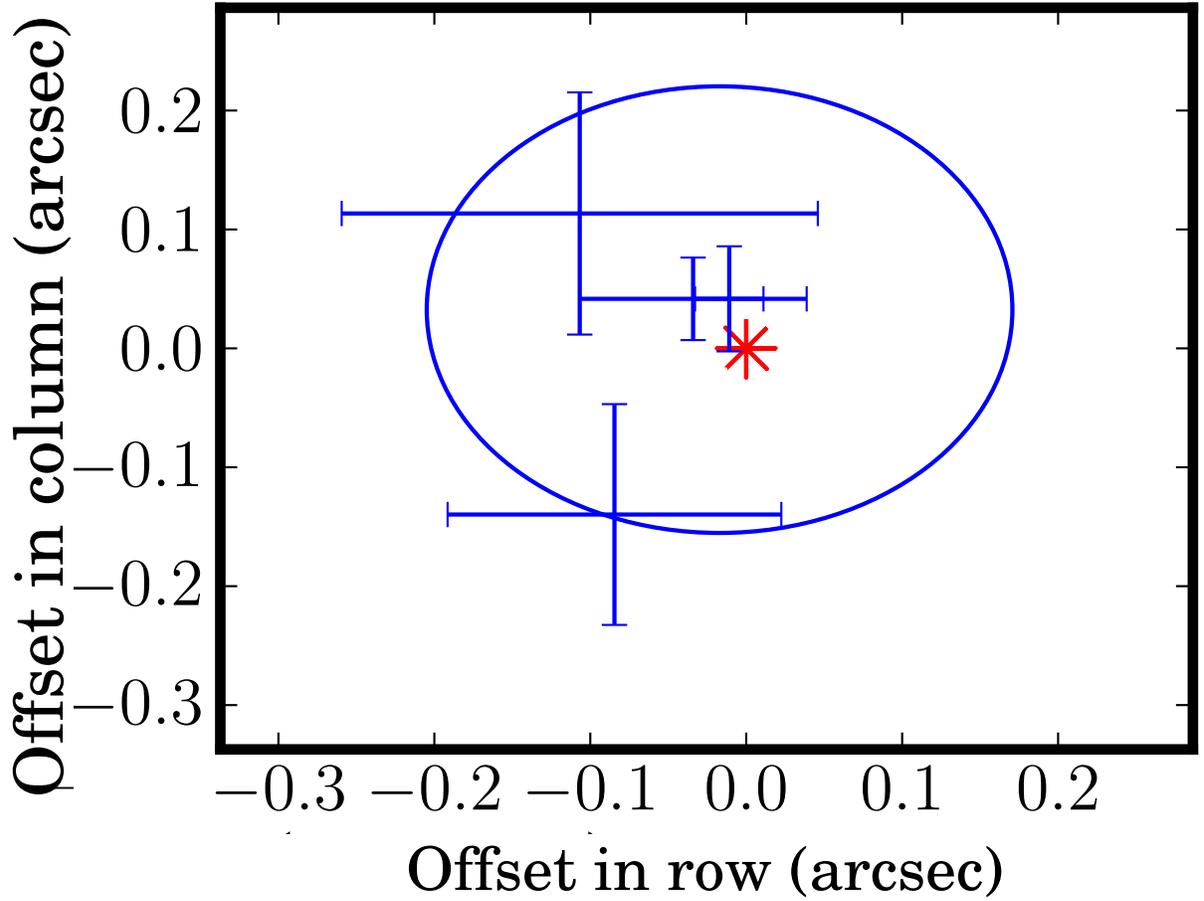}
\caption{Centoids offsets
for KIC~12735740. The red star shows the position of the star, the blue points are the
centroid position in transit relative to out of transit with associated uncertainty. The
blue circle is the 3-$\sigma$ uncertainty radius of confusion. All centroids are
consistent with the transit being on the target, although we cannot exclude false positives 
from sources within the blue circle.}
\label{fig:centroids}
\end{figure}

We carried out the false positive tests described in \S \ref{sec:FalsePosTest} for 
KIC~12735740 and place tight constraints on the possibility of contamination from 
a background or foreground star.
In Figure~\ref{fig:centroids} we show the centroid offsets for KIC~12735740 in the
3-$\sigma$ confusion region. The difference in the flux center in and out of transit was
calculated by taking the weighted average of the individual measurements using the
variance to define the weights. The 3-$\sigma$ uncertainty radius, R, is defined by
\begin{equation} R = 3 \times \frac{ \sqrt{\sum\limits^{n}\sigma^{2}} }{n} \, .
\end{equation} The 3-$\sigma$ uncertainty radius is 0.20 arcsec for KIC~12735740.

To derive the lower limit in $\Delta K_{P}$ we assume the worst case scenario: that 
the transit is caused by a diluted total eclipse. With this assumption, the brightest 
the eclipsed star can be is
\begin{equation} \Delta K_{P} = -2.5\log{(2T_{d})} \end{equation} where $T_{d}$ is the
observed transit depth. For KIC~12735740 we determine $\Delta K_{P}$ cutoff to be 5.01.

We can also place some constraints on the presence of a background/foreground star with 
our SME analysis. If two blended stars have nearly the same radial velocity, the spectral 
lines will be perfectly aligned; this scenario of physically aligned stars with the 
same radial velocity would be difficult to exclude, however it is also highly unlikely 
that two stars with chance alignment will also have the same radial 
velocity. We expect that a contaminating source is more likely to either broaden the 
observed spectral lines or to produce well separated lines. However, the spectral lines 
for KIC~12735740 are narrow (consistent with low vsini) and we do not see any evidence 
for extra flux at the blue or red edges of the spectral lines. In addition, at the level 
of the noise in our spectra we do not see any evidence for a second set of well-separated 
spectral lines. Unless nature has conspired against us so that the radial velocities are 
nearly identical, the flux contribution of any background/forground star is less 
than signal-to-noise ratio of 50 in our Keck/HIRES spectra for KIC~12735740. This constrains a 
putative interloper to have a magnitude difference that is $\ge 4.2$ magnitudes. 

If there is a faint binary companion and we reason that this companion star is on the 
same isochrone as KIC~12735740, the limit on stellar magnitude allows us to infer a maximum mass 
of 0.67 $M_{\odot}$ using a Dartmouth 4 Gyr isochrone~\citep{Dotter2008} and 
suggests that the putative interloper is a late K or M dwarf star with $T_{eff} \sim 4300$~K. 
Only three stars cooler than 4300 K in the $Kepler$ planet candidates 
list host gas giant ($>6.0 R_{\oplus}$) transiting planet candidates \citep{Batalha2012}. 
Qualitatively, this makes the scenario of a transiting planet 
around a low mass binary companion unlikely. Furthermore, a
dilution of 90\% would result in a transit depth of 0.098, which equates 
to an approximate scaled planet radius of $R_{p}/R_{\star}=0.3$. This results 
in a planet size that is not sub-stellar. Hence, we dismiss this scenario. 
We can also use this reasoning to dismiss the (unlikely) scenario that the 
transit is owing to a hierarchical stellar triple because
we would detect a secondary eclipse.

\begin{figure}
\includegraphics[width=\textwidth]{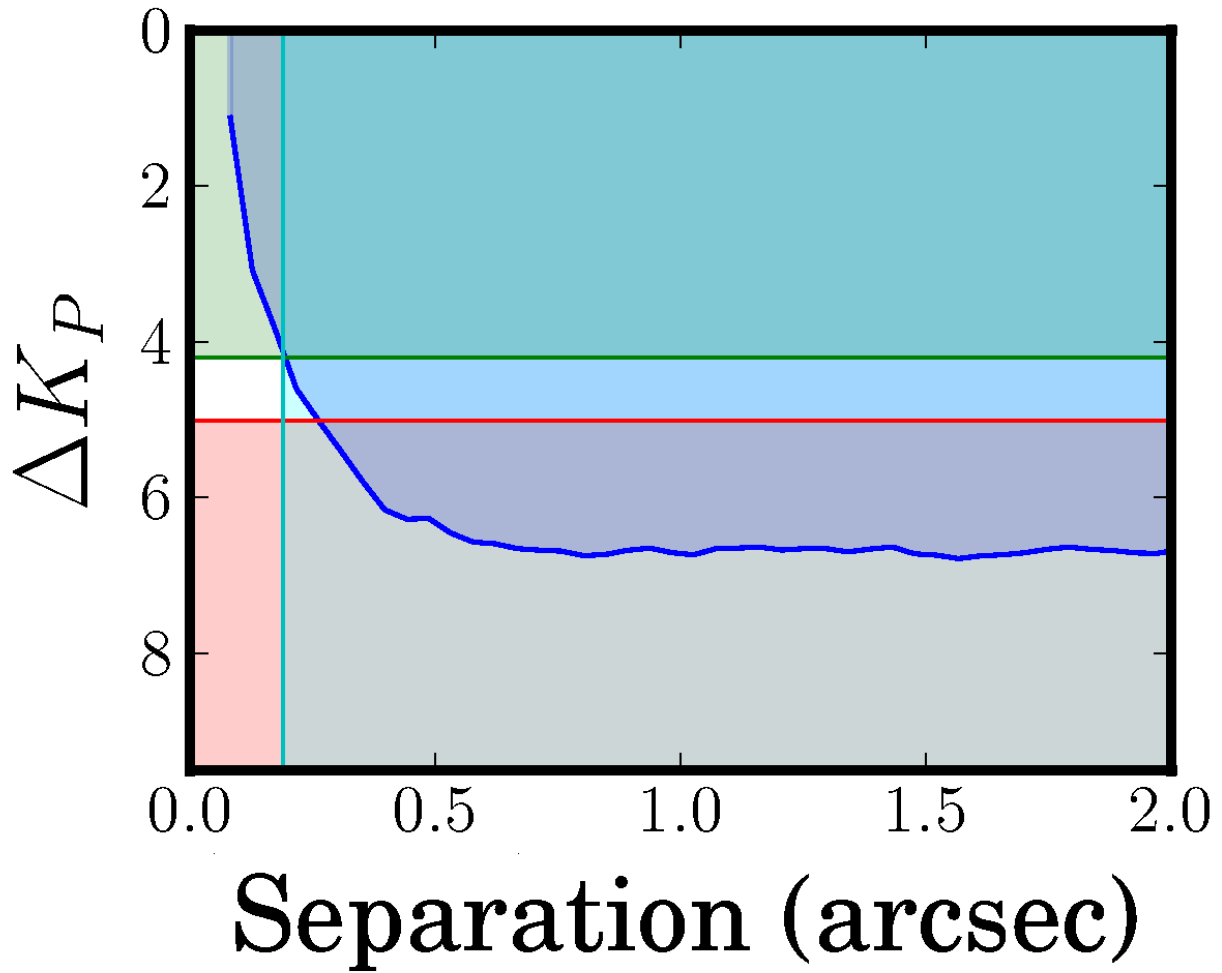}
\caption{Regions excluded
from our false positive calculations for KIC~12735740. We exclude parameter space using
spectroscopic inspection (green), AO imaging (blue), the transit curve analysis (red)
and pixel centroid analysis (cyan). Only the remaining white region is not excluded and 
we determine our false positive probability by estimating how many stars 
are statistically likely to appear in this space.}
\label{fig:exclusion}
\end{figure}

Figure~\ref{fig:exclusion} shows the regions of parameter space we can exclude. For
KIC~12735740, the number of background or foreground stars in the galaxy that might 
reside in the non-excluded space is statistically 0.000035. 
However, not all these background stars are likely to be eclipsing binaries or 
bright stars with transiting planets. According to Equation \ref{eq:pc}, our planet confidence
for KIC~12735740 is therefore 99.92\%. 

We obtained 4 RV data points for KIC~12735740 using Keck HIRES from UT 2013 June 3rd to 2013 June 25th, spanning 22 days. The results are reported in Table \ref{tab:RV_params}. The uncertainties of RV measurements are $\sim$ 2 m $\rm{s}^{-1}$. Neither periodical signal nor linear trend is detected, the rms of measurements is 14.0 m $\rm{s}^{-1}$. The rms indicates that KIC~12735740 has a large RV jitter level at $\sim$15 m $\rm{s}^{-1}$. We conducted Monte Carlo simulations to test the scenario of the transiting object being an EB. In simulations, we adopted the orbital solution obtained in this study (Table \ref{tab:orbital_params}). The argument of pariastron is randomized with a uniform distribution between 0 and 2$\pi$ because it is poorly constrained. The mass of the companion is assumed to be 80 $M_J$, a rough dividing line between a low-mass star and a Brown Dwarf (BD). We compared the range of the simulated RV curve and the rms of the measurements for the same time span as the HIRES observation. If the RV range is three times more than the rms, then the EB scenario is excluded at 3-sigma significance. We found the EB scenario is excluded in 95.7\% of simulations, implying that such possibility is low. We also considered zero eccentricity in simulations, and the possibility of the EB scenario is even lower. We also considered companions of lower mass, e.g., BD or planet scenario. The results are reported in Table \ref{tab:simulation_params}. We found a large fraction of simulations in which the BD scenario is excluded, although the fraction is smaller than that for the EB scenario. At current rms level, no strong constraint can be set on the planet scenario. 

Stellar companions are found around 46\% of solar-type stars~\citep{Raghavan2010}, and 2.8\% of the companions (M6V-M9V) have similar radii to the Jupiter, confusing the nature of PH2 b. However, the stellar companion (EB) scenario is excluded at high confidence by the Keck HIRES observations. BD companions are very rare compared to stellar companions and giant planets~\citep{Grether2006}. The occurrence rate is less than 1\% for BDs~\citep{Marcy2000}, 1.3\% (46\%$\times$2.8\%) for M6V-M9V companions~\citep{Duquennoy1991, Raghavan2010} and 5\% for planets~\citep{Fischer2005,Udry2007,Johnson2007}. In addition, BDs with mass of more than 20 $M_J$ is excluded by the HIRES observations at a relatively high confidence. Given the $\sim$1:50 relative probability of a BD to a giant planet~\citep{Grether2006} and the high plaent confidence, PH2 b should be a giant planet with a very high likelihood. 
 
\subsection{KIC~4820550}
The $Kepler$ magnitude for KIC~4820550 is 13.9. The star exhibits modest 
variability of $\sim$2\% in the $Kepler$ light curve data. The isochrone 
analysis suggests a stellar radius of $0.77 \rm{R}_\odot$. We compared the 
Ca II line core emission in the spectra of KIC~4820550 to CPS stars with 
similar T$_{\rm{eff}}$ of 5682~K and found that the Ca II H line core 
emission for HD~103432, a star with moderate chromospheric activity and \rhk\ 
of -4.63, was an excellent match. Five transits for KIC~4820550 were 
identified by Planet Hunters beginning with MJD 55124.54794, yielding an orbital 
period of 202.1175$\pm$0.0009~d. The radius of this planet candidates 
is modeled to be $5.13\pm0.32$ R$_\oplus$ and the planet confidence level was 
assessed at 88\%.

\subsection{KIC~4947556}
According to the $Kepler$ Input Catalog, KIC~4947556 has a $Kepler$ magnitude of 13.336 
and g-r color of 0.806. We derived $\rm{T}_{\rm{eff}}$ = $5122^{+95}_{-96}$~K, log $g$ =
$4.655^{+0.022}_{-0.032}$ and [Fe/H] = $+0.258^{+0.067}_{-0.049}$ for KIC~4947556 using 
SME analysis of the Keck HIRES spectrum. Our iterative isochrone analysis yields 
a stellar radius of $0.73^{+0.01}_{-0.02}$ $\rm{R}_\odot$. 
The emission of Ca H\&K line is comparable to the chromospherically active K dwarf, 
Epsilon Eridani (HD~22049), which has a chromospheric activity indicator \rhk$= -4.44$. 
The photometry of KIC 4947556 shows $\sim$6\% variability from star spots. 
Against the backdrop of this light curve variability, the Planet Hunters were able to 
detect transit signals with a depth of about 0.13\% at a period
of $140.6091\pm0.0045$~d.

Iterative analysis indicates a radius of 2.60$\pm$0.08 R$_\oplus$, 
making this a SuperEarth or mini-Neptune candidate in the habitable zone 
with T$_{\rm{PL}}=277\pm6$~K. This is the smallest planet candidate reported in this 
paper. Unfortunately, it would be challenging to obtain radial
velocity followup to determine the planet candidate mass and composition because 
of the faintness of the star and the small reflex velocity expected for 
the 141~d orbital period of this planet candidate. \citet{Tenenbaum2012} reported 
a TCE for this star with a 13.0268~d period and a transit depth of 425 ppm.

\subsection{KIC~9958387}
KIC~9958387 has a $Kepler$ magnitude of 13.5. Our spectroscopic analysis is consistent
with the KIC parameters of T$_{\rm{eff}}$=6159 K, log $g$=4.339 and [Fe/H]=-0.263. 
However, we derive a significantly different stellar radius. The KIC stellar radius 
is listed as 1.168 R$_\odot$ while our analysis indicates a larger radius of 
1.67$^{+0.13}_{-0.11}$ R$_\odot$. We estimate an upper limit for 
\rhk\ of -4.7 for KIC~9958387 by comparing it with spectra of CPS stars, so this is a 
relatively inactive star. Planet Hunter volunteers discovered five transits in the 
Q1 - 13 light curves, yielding an orbital period of $237.7886\pm0.0053$~d with the 
first detected transit at MJD of 55029.46560 day. 
Adopting our larger stellar radius of 1.67$^{+0.13}_{-0.11}$ R$_\odot$, the planet radius is 
estimated to be $8.78\pm0.81$ R$_\oplus$.

\subsection{Planet Candidates in Multi-planet systems}
\begin{figure}
\begin{center}
\includegraphics[width=16cm,height=16cm]{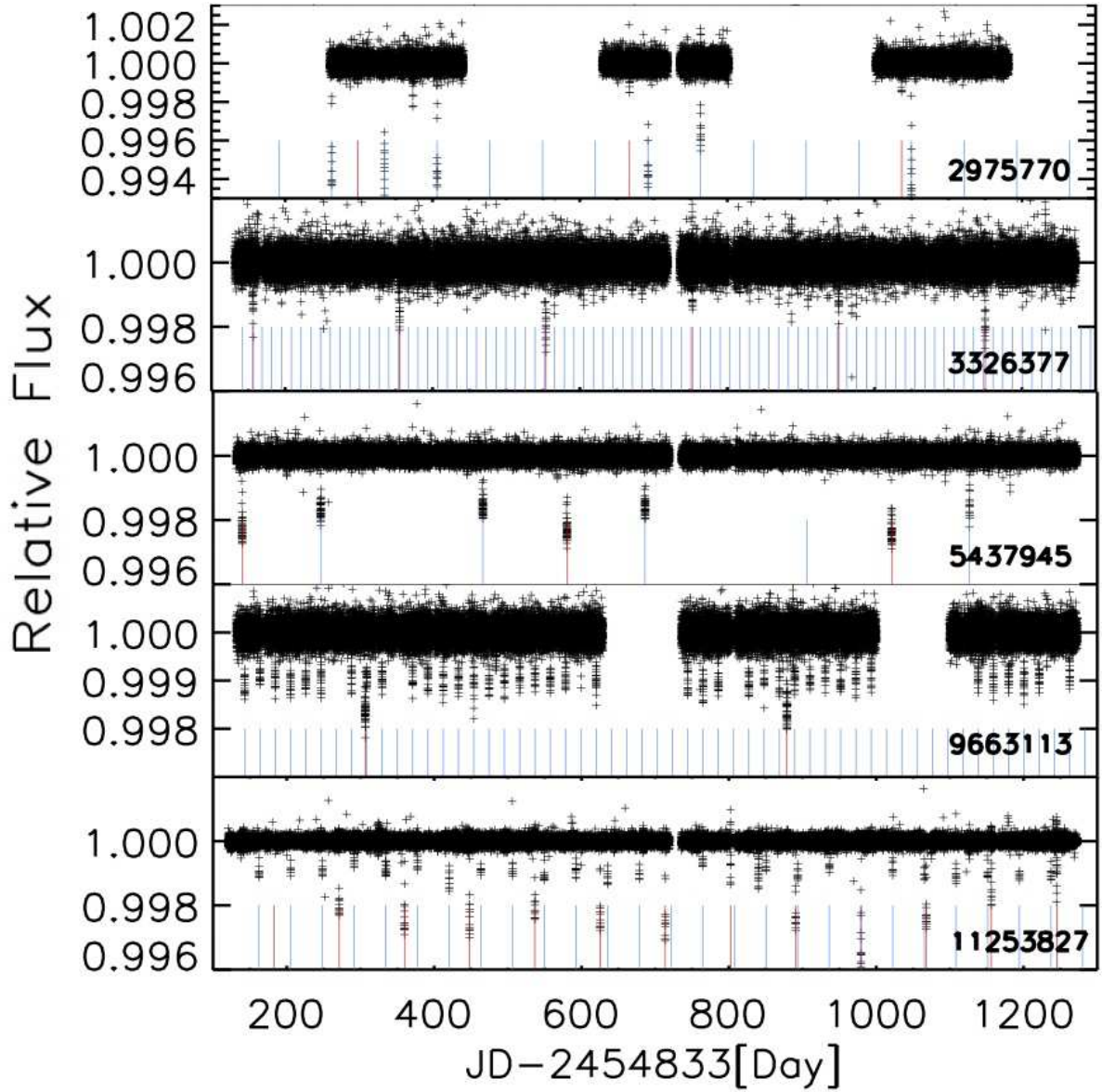} 
\caption{De-trended time series data for the multi-planet candidates, 
KIC~2975770, KIC~3326377, KIC~5437945, KIC~9663113 and KIC~11253827. 
Blue lines mark the transit times of the inner planet candidate,
red lines mark the transit times of the outer candidate.  
\label{fig:Mulptiple_LowRes}}
\end{center}
\end{figure}

Discoveries of transits of additional objects within the same system reduces the
probability of false positive because it is rare to have multiple false positive 
signals that mimic transiting planets. The detection of of two planet candidates around 
a single star increases the likelihood that a transit signal is a planet by a factor of 
$\sim$25 \citep{Lissauer2011}. \citet{Lissauer2012} further pointed out that 
almost all $Kepler$ multiple-planet candidates are true planets. The Planet Hunter 
volunteers discovered seven new multi-planet candidates; three of these were second 
planet candidates in systems that had already been announced by the $Kepler$ team. 
Figure \ref{fig:Mulptiple_LowRes} shows five multiple-planet candidate systems with both planets discovered by the Planet Hunters or the second candidate discovered by the Planet Hunters in addition to a KOI  (Kepler Object of Interest). 

KIC~5437945 has an effective temperature of 6093~K according to the KIC. 
Planet Hunter volunteers identified a pair of planet candidates in a 2:1 commensurability 
with orbital periods of 220.1344~d and 440.7704~d. In this case, the period ratio of 
2.003 further increases confidence for the planet interpretation for the transit 
light curve. Both planet candidates have Jupiter-like radii consistent with the 
interpretation that these are prospective gas giant planets. This pair of planets were also independently found by~\citet{Huang2012}. 

KIC~11253827 is the brightest star among our multi-planet candidate systems, with 
a $Kepler$ magnitude of 11.92. We obtained a Keck HIRES spectrum to derive 
the stellar properties and find that this is a sunlike star. With our stellar 
radius measurement (R$_\ast$ = 0.90$^{+0.06}_{-0.05}$ R$_\odot$), we derive radii for 
the two transiting planet candidates of $2.63\pm0.24$ and $4.42\pm0.32$ R$_\oplus$ 
respectively. The orbital periods are 42.99d and 88.5d; a period ratio of 2.06.~\citet{Huang2012} independently found this pair of candidates.

Planet Hunters also identified transiting planet candidates around three 
stars that were already known to have one other transit candidate (KOI~1788, 
KOI~179 and KOI~1830). A second planet candidate with a period of 369.1~d 
(nearly one Earth year) was discovered around KIC~2975770 (KOI~1788). 
The first planet candidate was announced by the $Kepler$ team with an 
orbital period of 71.5~d and a radius estimate of ($R_{P}=5.44\pm$0.31R$_\oplus$) 
\citep{Batalha2012}. The new candidate, KIC~2975770~c, appears to similar 
in size to Neptune with 4.22$\pm$2.50 R$_\oplus$. The equilibrium temperature for 
this outer planet is 243$\pm$44 K. The configuration of the two candidates, 
KIC~2975770 b and c, is close to a 5:1 mean motion resonance. 

The $Kepler$ team discovered a planet candidate with 3.07$\pm$0.14 R$_\oplus$ in a
20.7~d orbit around KOI~179 system (KIC~9663113) \citep{Batalha2012} and 
Planet Hunter volunteers discovered two transits that appear to arise from a 
second candidate with 10.93$\pm$1.38 R$_\oplus$ in a 572.4~d orbit. 
\citet{Ofir2012} announced a single transit for this outer planet, however 
their estimated planet radius is about half of our value. 

Planet Hunters found a more distant transit candidate with a radius of 
4.25$\pm$1.03 R$_\oplus$ in a 198.7~d orbit orbiting KIC~3326377 (KOI~1830).  
The $Kepler$ team had announced a super-Earth candidate with 1.76$\pm$0.09 R$_\oplus$ 
in a 13.2~d orbit about this same star. The outer planet in this system with a 198.7~d orbit was also announced by~\citet{Huang2012}. 

A pair of planets initially on coplanar circular orbits will be stable if they do not
develop crossing orbits, which will be true when the following condition is
met \citep{Lissauer2011}: 
\begin{equation} 
\Delta=\frac{a_2-a_1}{R_H}>2\sqrt{3},
\end{equation} 
where $R_H$ is the mutual Hill sphere and ($a_2 - a_1$) is the difference in 
the semi-major axes for the two planets. We find that this condition is strongly 
met for the 5 multi-planet systems presented here: KIC~2975770 ($\Delta = 21.2$), 
KIC~3326377 ($\Delta = 43.7$), KIC~9663113 ($\Delta = 32.8$), KIC~5437945 ($\Delta = 7.1$),
and KIC~11253827 ($\Delta = 15.6$).

According to a recent paper by~\citet{Tenenbaum2012}, KIC~4947556 and KIC~6878240 also
have TCEs with different periods than were found here. These may represent two 
more multi-planet candidate systems. 


\section{Summary and Discussion}
\label{sec:Summary}
In this paper, we provide a progress report of 43 new discoveries emerging 
from Planet Hunters. We have developed an iterative algorithm to combine
stellar evolution model and transiting light curve analysis to
obtain self-consistent stellar and orbital properties and model these systems 
(Table \ref{tab:orbital_params}) and carry out a false positive check for 
these candidates. 

The remaining 
planet candidates have three or more identified transits and most have orbital 
periods that exceed 100 days. This contribution increases by $\sim$30\% the sample of 
gas giant planet candidates \citep{Batalha2012} with orbital periods longer than 100~d 
and with radii that range between Neptune and Jupiter. 

Among these new candidates, twenty appear to orbit 
at distances where the temperature at the top of the atmosphere would be consistent
with temperatures in habitable zones. Most of these habitable zone planet candidates 
have radii comparable to or larger than Neptune; however, one candidate (KIC~4947556) 
has a radius of 2.60$\pm$0.08 R$_\oplus$ and may be a SuperEarth or mini-Neptune. 

At least nine of our planet candidates are members of (seven) multi-planet systems. 
The detection of multiple transit signals in a given light curve raises the confidence
level that prospective transits are truly planets \citep{Lissauer2012}.  
We were able to place severe limits on false positive probabilities for one 
of our candidates with follow-up spectroscopic analysis and AO imaging, 
pixel centroid analysis, galactic population synthesis and Bayesian planet priors. 
The planet confidence level for KIC~12735740 is 99.92\%, high enough that this 
seems to be a \textit{bona fide} planet and we assign the designation PH2~b for 
this planet. PH2~b is a Jupiter-size planet orbiting in the habitable zone of 
KIC~12735740, a solar-type star. 

Despite of residing in the habitable zone, PH2~b itself may not be habitable because of its gas giant planetary nature. However, exo-moons around PH2~b may be suitable for life to survive and evolve~\citep{Heller2013} and such exo-moons are not rare according to~\citet{Sasaki2010} in a study on the formation of the Jupiter and the Saturn satellite systems. What is more, ~\citet{Kipping2009a,Kipping2009b} has shown that an exo-moon with a mass of more than 20\% of Earth mass can be detected in the $Kepler$ mission. And the first result of searching for exo-moons is reported in ~\citet{Kipping2013}. As the time baseline of the $Kepler$ mission increases, PH2~b and other candidates reported in this paper may become suitable for exo-moon search in the near future. 

\begin{figure}
\begin{center}
\includegraphics[width=16cm,height=11cm]{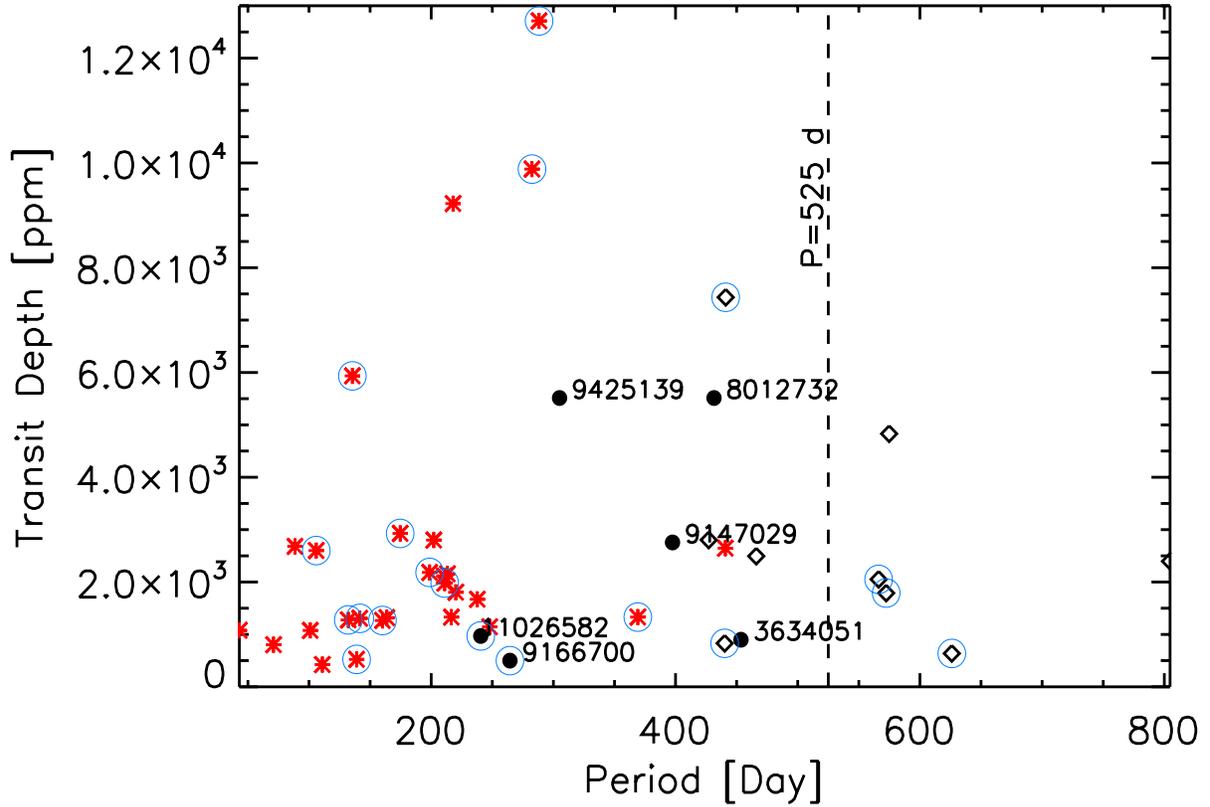} 
\caption{Transit depth
and orbital period scatter plot for 43 planet candidates presented in this paper. Red
asterisks are candidates also included by the $Kepler$ TCEs list \citep{Tenenbaum2012}. 
Black solid circles marked with KIC numbers are discoveries not included on the TCEs list
but with at least 3 transits identified by Planet Hunter participants.
Diamonds are long period candidates where only two transits were found by 
Planet Hunter volunteers. Planet candidates that appear to reside in the habitable zone 
are circled in blue. 
\label{fig:Depth_vs_Period_PH_TCE}}
\end{center}
\end{figure}

~\citet{Tenenbaum2012} recently presented detections of potential transit signals in the
first twelve quarters of $Kepler$ mission data and provided a list of TCEs (Threshold
Crossing Events). Figure \ref{fig:Depth_vs_Period_PH_TCE} plots the  
planet candidates presented in this paper that are also on the TCEs list. 
Among our 43 new candidates, 28 are included in the TCEs list with orbital parameters 
that agree and 15 are unique findings. 

Planet Hunters found 6 additional planet candidates with periods shorter than 525~d 
(the boundary imposed by the requirement that at least three transits must be detected by 
the Kepler TCE pipeline) that have three or more transits: KIC~3634051, KIC~8012732, 
KIC~9147029, KIC~9166700, KIC~9425139 and KIC~11026582. 
The other nine of the unique candidates have prospective periods longer than 400~d
so only two transits were found in the Q1 - 13 data; therefore, these would not have 
been detected by the TCE pipeline.  
However, the estimated orbital periods for these nine candidates allow us to establish 
a watch list on the Planet Hunters site to intensify the search for subsequent transits.

\noindent{\it Acknowledgements} 

The data presented in this paper are the result of the efforts of the Planet Hunters volunteers, without whom this work would not have been possible. Their contributions are individually acknowledged at \url{http://www.planethunters.org/authors}. The authors thank the  Planet Hunters volunteers who participated in identifying and analyzing the candidates presented in this paper. They are individually recognized at \url{http://www.planethunters.org/LongPeriodCandidates}

DF acknowledges funding support for PlanetHunters.org from Yale University and support 
from the NASA Supplemental Outreach Award, 10-OUTRCH.210-0001 and the NASA ADAP12-0172. 
MES is supported by an NSF Astronomy and
Astrophysics Postdoctoral Fellowship under award AST-100325 and in part by an American
Philosophical Society Grant. KS gratefully acknowledges support from Swiss National
Science Foundation Grant PP00P2\_138979/1. The Zooniverse is supported by The Leverhulme
Trust and by the Alfred P. Sloan foundation. Planet Hunters is supported in part by 
NASA JPL's PlanetQuest program. 
The Talk system used by Planet Hunters was built during work supported by the
National Science Foundation under Grant No. DRL-0941610. We gratefully acknowledge the
dedication and achievements of $Kepler$ science team and all those who contributed to the
success of the mission. We acknowledge use of public release data served by the
NASA/IPAC/NExScI Star and Exoplanet Database, which is operated by the Jet Propulsion
Laboratory, California Institute of Technology, under contract with the National
Aeronautics and Space Administration. This research has made use of NASAÕs Astrophysics
Data System Bibliographic Services. This paper includes data collected by the $Kepler$
spacecraft, and we gratefully acknowledge the entire $Kepler$ mission teamÕs efforts in
obtaining and providing the light curves used in this analysis. Funding for the $Kepler$
mission is provided by the NASA Science Mission directorate. The publicly released
$Kepler$ light curves were obtained from the Multimission Archive at the Space Telescope
Science Institute (MAST). STScI is operated by the Association of Universities for
Research in Astronomy, Inc., under NASA contract NAS5-26555. Support for MAST for non-HST
data is provided by the NASA Office of Space Science via grant NNX09AF08G and by other
grants and contracts.

\bibliography{mybib_JW_DF_PH5}

\clearpage


\newpage

\clearpage

\begin{deluxetable}{lcccccc}
\tablewidth{0pt}
\tablecaption{$\Delta K$ magnitude AO exclusion limits.\label{tab:AO_params}}
\tablehead{
\colhead{\textbf{Star Name}} &
\colhead{\textbf{0.25}} &
\colhead{\textbf{0.50}} &
\colhead{\textbf{1.00}} &
\colhead{\textbf{1.50}} &
\colhead{\textbf{2.00}} &
\colhead{\textbf{2.50}} \\
\colhead{\textbf{KIC}} &
\colhead{\textbf{(arcsec)}} &
\colhead{\textbf{(arcsec)}} &
\colhead{\textbf{(arcsec)}} &
\colhead{\textbf{(arcsec)}} &
\colhead{\textbf{(arcsec)}} &
\colhead{\textbf{(arcsec)}} \\

}

\startdata

4947556 & 2.48 & 4.17 & 4.91 & 4.84 & 4.85 & 4.79 \\
4820550 & 2.87 & 3.96 & 4.15 & 4.12 & 4.09 & 4.10 \\
9958387 & 4.67 & 5.62 & 5.56 & 5.57 & 5.62 & 5.54 \\
12735740 & 4.91 & 6.32 & 6.72 & 6.73 & 6.68 & 6.63 \\

\enddata

\end{deluxetable}

\clearpage

\newpage

\clearpage

\begin{deluxetable}{lccccc}
\tablewidth{0pt}

\tablecaption{SME results.\label{tab:SME_params}}
\tablehead{
\colhead{\textbf{Star Name}} &
\colhead{\textbf{$T_{\rm{eff}}$}} &
\colhead{\textbf{log $g$}} &
\colhead{\textbf{[Fe/H]}} &
\colhead{\textbf{[$\alpha$/Fe]}} &
\colhead{\textbf{V$\sin i$}} \\
\colhead{\textbf{KIC}} &
\colhead{\textbf{(K)}} &
\colhead{\textbf{(cgs)}} &
\colhead{\textbf{(dex)}} &
\colhead{\textbf{(dex)}} &
\colhead{\textbf{(km s$^{-1}$)}} \\

}

\startdata

        KIC2581316 &
$6148\pm132$ &
$ 4.170\pm 0.180$ &
$ 0.111\pm 0.090$ &
$ 0.033\pm 0.174$ &
$ 7.35\pm 1.50$ \\
        KIC4472818 &
$5848\pm88$ &
$ 4.480\pm 0.120$ &
$ 0.119\pm 0.060$ &
$-0.035\pm 0.116$ &
$ 2.64\pm 1.44$ \\
        KIC4820550 &
$5678\pm50$ &
$ 4.686\pm 0.060$ &
$ 0.128\pm 0.033$ &
$ 0.021\pm 0.075$ &
$ 0.95\pm 0.95$ \\
        KIC4947556 &
$5110\pm88$ &
$ 4.800\pm 0.120$ &
$ 0.260\pm 0.060$ &
$-0.113\pm 0.144$ &
$ 5.80\pm 1.00$ \\
        KIC9147029 &
$5868\pm64$ &
$ 4.551\pm 0.248$ &
$ 0.051\pm 0.049$ &
$ 0.090\pm 0.330$ &
$ 0.32\pm 0.50$ \\
        KIC9958387 &
$6133\pm44$ &
$ 4.078\pm 0.060$ &
$ 0.086\pm 0.030$ &
$ 0.034\pm 0.058$ &
$ 5.22\pm 0.53$ \\
       KIC11253827 &
$5666\pm88$ &
$ 4.594\pm 0.120$ &
$ 0.171\pm 0.070$ &
$-0.022\pm 0.132$ &
$ 3.31\pm 1.00$ \\
       KIC11392618 &
$5708\pm65$ &
$ 4.288\pm 0.088$ &
$ 0.011\pm 0.039$ &
$ 0.094\pm 0.078$ &
$ 1.62\pm 1.04$ \\
       KIC12735740 &
$5631\pm44$ &
$ 4.418\pm 0.060$ &
$-0.079\pm 0.030$ &
$ 0.020\pm 0.062$ &
$ 1.43\pm 0.78$ \\

\enddata

\end{deluxetable}

\clearpage

\newpage

\clearpage

\begin{landscape}
\begin{deluxetable}{lcccccccccccccccc}
\tabletypesize{\tiny}
 \setlength{\tabcolsep}{0.02in} 
\tablewidth{0pt}
\tablecaption{Stellar Parameters\label{tab:stellar_params}}
\tablehead{
\colhead{\textbf{ }} &
\multicolumn{7}{c}{\textbf{KIC}} &
\multicolumn{9}{c}{\textbf{Derived Values}} \\
\colhead{\textbf{Star}} &
\colhead{\textbf{$\alpha$}} &
\colhead{\textbf{$\delta$}} &
\colhead{\textbf{$Kp$$^{\dagger}$}} &
\colhead{\textbf{($g-r$)$^{\dagger}$}} &
\colhead{\textbf{$T_{\rm eff}$$^{\dagger}$}} &
\colhead{\textbf{[Fe/H]$^{\dagger}$}} &
\colhead{\textbf{$\log g$$^{\dagger}$}} &
\colhead{\textbf{$T_{\rm eff}$}} &
\colhead{\textbf{$\log g$}} 	&
\colhead{\textbf{M$_{\ast}$}} &
\colhead{\textbf{R$_{\ast}$}}	&
\colhead{\textbf{L$_{\ast}$}}	&
\colhead{\textbf{$\rho_{\ast}$}}	&
\colhead{\textbf{[Fe/H]}}	&
\colhead{\textbf{v $\sin i$}}	&
\colhead{\textbf{$\delta f/f$}}	\\
\colhead{\textbf{(KIC)}} &
\colhead{\textbf{(h m s)}} &
\colhead{\textbf{(d m s)}} &
\colhead{\textbf{(mag)}}	&
\colhead{\textbf{(mag)}}	&
\colhead{\textbf{(K)}}	&
\colhead{\textbf{(dex)}}	&
\colhead{\textbf{(cgs)}}	&
\colhead{\textbf{(K)}}	&
\colhead{\textbf{(cgs)}}	&
\colhead{\textbf{(M$_{\odot}$)}} &
\colhead{\textbf{(R$_{\odot}$)}} &
\colhead{\textbf{(L$_{\odot}$)}} &
\colhead{\textbf{(g~cm$^{-3}$)}} &
\colhead{\textbf{(dex)}} &
\colhead{\textbf{(km~s$^{-1}$)}} &
\colhead{\textbf{(\nodata)}}
}

\startdata

2581316$^{\dagger\dagger}$     &   19 30 30.590&   +37 51 36.51&    11.69    &    0.680    &    \nodata    &    \nodata    &    \nodata    & $    6147^{+     147}_{-     104}$ & $       3.90^{+       0.04}_{-       0.04}$ & $       1.29^{+       0.31}_{-       0.35}$ & $       2.10^{+       0.17}_{-       0.20}$ & $       2.84^{+       1.19}_{-       1.32}$ & $     0.19\pm     0.07$ & $       0.13^{+       0.04}_{-       0.10}$ & 7.35(1.50) &    2.00e-04 \\

2975770    &   19 10 57.014&   +38 06 52.42&    14.52    &    0.951    &    4683    &    $0.24$\phn    &    4.34    &    $4730  ^{+177   }_{ -189  }$    &    $4.52  ^{ +0.07  }_{ -0.82 }$    &    $0.87  ^{ +0.34  }_{ -0.11 }$    &    $0.82  ^{ +1.73  }_{ -0.08 }$    &    $0.35  ^{ +3.14  }_{ -0.12 }$    &    $2.19  \pm{ 7.27 }$    &    $0.25  ^{ +0.15  }_{ -0.27 }$    &    \nodata    &    7.44e-03 \\

3326377    &   19 06 13.939&   +38 24 19.33&    14.44    &    0.703    &    4926    &    $-0.86$\phn\phs    &    4.49    &    $4991  ^{ +197   }_{ -183  }$    &    $4.65  ^{ +0.3  }_{ -0.3 }$    &    $0.67  ^{ +0.06  }_{ -0.04 }$    &    $0.64  ^{ +0.06  }_{ -0.03 }$    &    $0.24  ^{ +0.08  }_{ -0.05 }$    &    $3.61  \pm{ 0.80 }$    &    $-0.78 ^{ +0.40  }_{ -0.02 }$    &    \nodata    &    6.62e-03 \\

3634051    &   19 11 54.110&   +38 44 25.01&    13.44    &    0.478    &    5831    &    $-0.08$\phn\phs    &    4.30    &    $5846^{+211}_{-198}$    &    $3.8^{+0.2}_{-0.2}$    &    $1.13^{+0.32}_{-0.14}$    &    $2.20^{+0.71}_{-0.75}$    &    $3.04^{+4.02}_{-1.48}$    &    $0.1\pm0.2$    &    $-0.1^{+0.4}_{-0.4}$    &    \nodata    &    2.39e-04 \\

3663173    &   19 42 40.594&   +38 47 45.02&    15.91    &    0.525    &    6116    &    $0.19$\phn    &    4.74    &    $6054^{+173}_{-150}$    &    $3.9^{+0.1}_{-0.2}$    &    $1.14^{+0.26}_{-0.12}$    &    $1.83^{+0.88}_{-0.21}$    &    $2.74^{+3.20}_{-1.02}$    &    $0.3\pm0.2$    &    $0.1^{+0.4}_{-0.4}$    &    \nodata    &    1.26e-03 \\

3732035    &   19 06 27.288&   +38 53 31.09&    14.21    &    0.464    &    5786    &    $-0.47$\phn\phs    &    4.59    &    $5791^{+193}_{-180}$    &    $3.9^{+0.2}_{-0.2}$    &    $1.09^{+0.27}_{-0.13}$    &    $1.94^{+0.80}_{-0.66}$    &    $3.69^{+3.33}_{-2.31}$    &    $0.2\pm0.2$    &    $-0.4^{+0.4}_{-0.5}$    &    \nodata    &    3.78e-04 \\

4142847    &   19 09 12.677&   +39 17 22.88&    15.20    &    0.823    &    4762    &    $-0.40$\phn\phs    &    4.59    &    $4759  ^{ +179   }_{ -175  }$    &    $4.63  ^{ +0.3  }_{ -0.3 }$    &    $0.69  ^{ +0.09  }_{ -0.06 }$    &    $0.67  ^{ +0.08  }_{ -0.06 }$    &    $0.20  ^{ +0.09  }_{ -0.05 }$    &    $3.29  \pm{ 1.13 }$    &    $-0.35 ^{ +0.36  }_{ -0.36 }$    &    \nodata    &    3.99e-03 \\

4472818$^{\dagger\dagger}$ &   19 35 32.496&   +39 31 11.32&    12.72    &    0.489    &    5650    &    $-0.35$\phn\phs    &    4.37    & $    5834^{+      96}_{-      75}$ & $       4.33^{+       0.05}_{-       0.07}$ & $       1.09^{+       0.05}_{-       0.09}$ & $       1.16^{+       0.09}_{-       0.08}$ & $       1.28^{+       0.31}_{-       0.20}$ &  $     0.97\pm      0.23$ & $       0.11^{+       0.06}_{-       0.05}$ & 2.64(1.44) &    5.53e-03 \\

4760478    &   19 41 43.044&   +39 53 11.54&    15.76    &    0.646    &    5430    &    $-0.27$\phn\phs    &    4.53    &    $5427^{+215}_{-202}$    &    
$3.72^{+0.20}_{-0.41}$&
$1.15^{+0.63}_{-0.20}$&
$2.39^{+2.21}_{-0.54}$&
$4.51^{+11.58}_{-1.94}$&
$0.12\pm0.2$    &
$-0.2^{+0.4}_{-0.6}$    &    \nodata    &    8.97e-04 \\

4820550$^{\dagger\dagger}$$^{,\dagger\dagger\dagger}$    &   19 07 05.146&   +39 59 01.75&    13.92    &    0.535    &    5586    &    $-0.15$\phn\phs    &    4.63    &    $5682^{+41}_{-46}$    &    $4.673^{+0.041}_{-0.044}$    &    $1.02^{+0.02}_{-0.03}$    &    $0.77^{+0.04}_{-0.03}$    &    $0.98^{+0.10}_{-0.13}$    &    $3.1\pm0.4$    &    $0.125^{+0.032}_{-0.028}$    &    0.95(0.95)    &    3.53e-03 \\

4902202    &   18 54 46.440&   +40 01 04.87&    15.58    &    0.595    &    5412    &    $-0.52$\phn\phs    &    4.49    &    $4926  ^{ +194   }_{ -200  }$    &    $4.52  ^{ +0.3  }_{ -0.3 }$    &    $0.87  ^{ +0.18  }_{ -0.08 }$    &    $0.83  ^{ +1.19  }_{ -0.07 }$    &    $0.40  ^{ +1.83  }_{ -0.12 }$    &    $2.12  \pm{ 4.85 }$    &    $0.24  ^{ +0.16  }_{ -0.27 }$    &    \nodata    &    9.52e-04 \\

4947556$^{\dagger\dagger}$$^{,\dagger\dagger\dagger}$    &   19 50 18.506&   +40 01 39.76&    13.34    &    0.806    &    4907    &    $0.22$\phn    &    4.51    &    $5122^{+95}_{-96}$    &    $4.655^{+0.022}_{-0.032}$    &    $0.89^{+0.03}_{-0.03}$    &    $0.73^{+0.01}_{-0.02}$    &    $0.43^{+0.07}_{-0.06}$    &    $3.0\pm0.2$    &    $0.258^{+0.067}_{-0.049}$    &    5.80(1.00)    &    1.08e-02 \\

5437945 &   19 13 53.962&   +40 39 04.90&    13.77    &    0.410    &    6093    &    $-0.37$\phn\phs    &    4.17    & $    6079^{+     201}_{-     188}$ & $       3.96^{+       0.10}_{-       0.31}$ & $       1.13^{+       0.23}_{-       0.15}$ & $       1.83^{+       0.93}_{-       0.27}$ & $       3.54^{+       3.43}_{-       1.57}$ &  $     0.25\pm      0.26$ & $      -0.39^{+       0.55}_{-       0.44}$ & \nodata &    2.44e-04 \\

5857656    &   18 58 38.453&   +41 06 32.18&    12.84    &    0.429    &    5937    &    $-0.06$\phn\phs    &    4.31    &    $5943^{+179}_{-217}$    &    
$3.95^{+0.24}_{-0.63}$&
$1.16^{+0.79}_{-0.18}$&
$1.87^{+3.36}_{-0.56}$&
$3.74^{+26.14}_{-2.10}$&
$0.25\pm0.78$    &
$-0.0^{+0.4}_{-0.4}$    &    \nodata    &    2.03e-04 \\

5871985    &   19 21 30.315&   +41 09 02.66&    15.17    &    1.058    &    4308    &    $-0.75$\phn\phs    &    4.54    &    $4360  ^{ +199   }_{ -185  }$    &    $4.69  ^{ +0.3  }_{ -0.3 }$    &    $0.58  ^{ +0.07  }_{ -0.04 }$    &    $0.57  ^{ +0.06  }_{ -0.04 }$    &    $0.11  ^{ +0.04  }_{ -0.03 }$    &    $4.39  \pm{ 1.27 }$    &    $-0.62 ^{ +0.44  }_{ -0.18 }$    &    \nodata    &    2.94e-03 \\

5966810    &   19 35 16.390&   +41 12 41.26&    14.95    &    0.414    &    6152    &    $-0.49$\phn\phs    &    4.63    &    $6149^{+174}_{-218}$    &    $3.8^{+0.1}_{-0.2}$    &    $1.13^{+0.27}_{-0.15}$    &    $2.02^{+0.67}_{-0.38}$    &    $4.01^{+3.14}_{-1.88}$    &    $0.2\pm0.2$    &    $-0.5^{+0.5}_{-0.5}$    &    \nodata    &    5.34e-04 \\

6106282    &   19 01 23.988&   +41 27 07.96&    15.13    &    1.368    &    3912    &    $-0.13$\phn\phs    &    4.41    &    $3963  ^{ +175   }_{ -169  }$    &    $4.69  ^{ +0.3  }_{ -0.3 }$    &    $0.60  ^{ +0.07  }_{ -0.06 }$    &    $0.58  ^{ +0.07  }_{ -0.05 }$    &    $0.08  ^{ +0.04  }_{ -0.02 }$    &    $4.29  \pm{ 1.42 }$    &    $0.05  ^{ +0.35  }_{ -0.38 }$    &    \nodata    &    2.78e-03 \\

6878240    &   19 43 34.586&   +42 22 49.62&    16.00    &    0.733    &    5144    &    $-0.04$\phn\phs    &    4.88    &    $5131^{+194}_{-193}$    &    $4.37^{+0.08}_{-0.10}$    &    $0.87^{+0.11}_{-0.10}$    &    $1.00^{+0.16}_{-0.13}$    &    $0.45^{+0.26}_{-0.16}$    &    $1.2\pm0.6$    &    $0.01^{+0.43}_{-0.47}$    &    \nodata    &    8.38e-04 \\

7826659    &   19 33 11.606&   +43 33 21.13&    13.86    &    0.969    &    4665    &    $0.48$\phn    &    4.10    &    $4675  ^{ +186   }_{ -189  }$    &    $4.58  ^{ +0.3  }_{ -0.3 }$    &    $0.82  ^{ +0.06  }_{ -0.05 }$    &    $0.78  ^{ +0.04  }_{ -0.04 }$    &    $0.29  ^{ +0.08  }_{ -0.07 }$    &    $2.47  \pm{ 0.42 }$    &    $0.40  ^{ +0.00  }_{ -0.09 }$    &    \nodata    &    1.34e-03 \\

8012732$^{,\dagger\dagger\dagger \dagger}$ &   18 58 55.079&   +43 51 51.18&    \nodata    &    0.511    &    \nodata    &    \nodata    &    \nodata    & $    6035^{+     437}_{-     389}$ & $       3.99^{+       0.06}_{-       0.37}$ & $       1.17^{+       0.25}_{-       0.12}$ & $       1.86^{+       1.03}_{-       0.20}$ & $       2.98^{+       2.41}_{-       1.07}$ &  $     0.25\pm      0.25$ & $      -0.06^{+       0.4}_{-       0.4}$ & \nodata &    3.09e-04 \\

8210018    &   18 42 02.822&   +44 09 33.66&    15.03    &    1.053    &    4337    &    $-0.72$\phn\phs    &    4.74    &    $4432  ^{ +186   }_{ -186  }$    &    $4.64  ^{ +0.3  }_{ -0.3 }$    &    $0.67  ^{ +0.06  }_{ -0.07 }$    &    $0.65  ^{ +0.05  }_{ -0.07 }$    &    $0.15  ^{ +0.05  }_{ -0.04 }$    &    $3.41  \pm{ 0.97 }$    &    $-0.20 ^{ 0.36  }_{ -0.42 }$    &    \nodata    &    2.07e-03 \\

8636333    &   19 43 47.585&   +44 45 11.23&    15.29    &    0.459    &    6034    &    $-0.30$\phn\phs    &    4.66    &    $6038^{+176}_{-187}$    &    $4.1^{+0.2}_{-0.3}$    &    $1.06^{+0.20}_{-0.14}$    &    $1.35^{+0.73}_{-0.31}$    &    $1.93^{+2.47}_{-0.82}$    &    $0.6\pm0.7$    &    $-0.2^{+0.4}_{-0.5}$    &    \nodata    &    6.40e-04 \\

8827930    &   19 40 49.663&   +45 05 53.48&    16.00    &    0.649    &    5581    &    $0.29$\phn    &    4.72    &    $5570^{+197}_{-207}$    &    $4.8^{+0.2}_{-0.3}$    &    $0.97^{+0.11}_{-0.10}$    &    $0.63^{+0.31}_{-0.17}$    &    $0.88^{+0.45}_{-0.30}$    &    $5.4\pm6.3$    &    $0.3^{+0.2}_{-0.5}$    &    \nodata    &    9.05e-04 \\

9025971 &   19 33 07.574&   +45 18 34.81&    14.52    &    0.524    &    5702    &    $-0.08$\phn\phs    &    4.71    & $    5695^{+     198}_{-     205}$ & $       4.61^{+       0.10}_{-       0.55}$ & $       1.00^{+       0.14}_{-       0.13}$ & $       0.80^{+       0.83}_{-       0.10}$ & $       1.12^{+       1.24}_{-       0.50}$ &  $      2.6\pm       4.6$ & $      -0.1^{+       0.4}_{-       0.4}$ & \nodata &    4.81e-04 \\

9147029$^{\dagger\dagger}$    &   19 13 44.594&   +45 31 11.46&    15.31    &    0.437    &    6216    &    $0.19$\phn    &    4.64    &    $5867^{+43}_{-42}$    &    $4.541^{+0.040}_{-0.039}$    &    $1.04^{+0.02}_{-0.04}$    &    $0.90^{+0.04}_{-0.04}$    &    $1.23^{+0.14}_{-0.16}$    &    $1.9\pm0.2$    &    $0.051^{+0.027}_{-0.031}$    &    0.32(0.50)    &    5.64e-04 \\

9166700    &   19 45 21.149&   +45 32 42.40&    11.15    &    0.311    &    6307    &    $-0.11$\phn\phs    &    4.00    &    $6318^{+186}_{-189}$    &    $4.2^{+0.2}_{-0.2}$    &    $1.08^{+0.20}_{-0.12}$    &    $1.35^{+0.56}_{-0.40}$    &    $2.12^{+1.33}_{-0.49}$    &    $0.6\pm0.7$    &    $-0.1^{+0.4}_{-0.5}$    &    \nodata    &    1.79e-04 \\

9413313    &   19 41 40.915&   +45 54 12.56&    14.12    &    0.713    &    5165    &    $-0.01$\phn\phs    &    4.26    &    $5169^{+221}_{-206}$    &    $4.3^{+0.3}_{-0.3}$    &    $0.92^{+0.20}_{-0.13}$    &    $1.03^{+0.67}_{-0.31}$    &    $0.61^{+1.86}_{-0.28}$    &    $1.2\pm1.7$    &    $-0.0^{+0.4}_{-0.4}$    &    \nodata    &    7.92e-03 \\

9425139 &   19 55 58.822&   +45 54 34.24&    13.38    &    0.635    &    5321    &    $-0.09$\phn\phs    &    4.26    & $    5330^{+     169}_{-     184}$ & $       3.98^{+       0.54}_{-       0.20}$ & $       1.07^{+       0.34}_{-       0.20}$ & $       1.84^{+       0.65}_{-       1.01}$ & $       2.3^{+       3.3}_{-       1.85}$ &  $    0.27\pm      0.32$ & $      -0.1^{+       0.4}_{-       0.4}$ & \nodata &    2.50e-04 \\

9480535    &   19 49 39.461&   +46 03 38.95&    15.13    &    0.520    &    5825    &    $-0.27$\phn\phs    &    4.52    &    $5847^{+209}_{-218}$    &    $3.95^{+0.18}_{-0.19}$    &    $1.15^{+0.24}_{-0.14}$    &    $1.90^{+0.65}_{-0.44}$    &    $3.71^{+2.65}_{-2.07}$    &    $0.2\pm0.2$    &    $-0.26^{+0.45}_{-0.46}$    &    \nodata    &    4.86e-04 \\

9663113    &   19 48 10.901&   +46 19 43.32&    13.96    &    0.483    &    5827    &    $-0.29$\phn\phs    &    4.42    &    $5816^{+230}_{-141}$    &    $3.8^{+0.1}_{-0.1}$    &    $1.21^{+0.39}_{-0.19}$    &    $2.33^{+0.30}_{-0.26}$    &    $4.56^{+5.07}_{-2.28}$    &    $0.1\pm0.1$    &    $-0.2^{+0.4}_{-0.3}$    &    \nodata    &    2.43e-04 \\

9886255    &   19 19 28.714&   +46 43 46.96&    15.80    &    0.730    &    5137    &    $0.01$\phn    &    4.83    &    $5115^{+225}_{-196}$    &    $4.31^{+0.17}_{-0.11}$    &    $0.88^{+0.10}_{-0.09}$    &    $1.06^{+0.18}_{-0.19}$    &    $0.51^{+0.32}_{-0.16}$    &    $1.1\pm0.6$    &    $0.06^{+0.44}_{-0.54}$    &    \nodata    &    1.85e-03 \\

9958387$^{\dagger\dagger}$$^{,\dagger\dagger\dagger}$    &   19 39 01.848&   +46 49 52.03&    13.46    &    0.387    &    6159    &    $-0.26$\phn\phs    &    4.36    &    $6137^{+40}_{-46}$    &    $4.068^{+0.038}_{-0.044}$    &    $1.21^{+0.14}_{-0.16}$    &    $1.67^{+0.13}_{-0.11}$    &    $2.64^{+1.45}_{-0.89}$    &    $0.32\pm0.1$    &    $0.085^{+0.029}_{-0.030}$    &    5.22(0.53)    &    2.00e-04 \\

10024862    &   19 47 12.602&   +46 56 04.42&    15.88    &    0.434    &    6412    &    $0.07$\phn    &    4.61    &    $6403^{+218}_{-213}$    &    $4.1^{+0.2}_{-0.2}$    &    $1.11^{+0.25}_{-0.12}$    &    $1.52^{+0.55}_{-0.31}$    &    $2.38^{+2.07}_{-0.61}$    &    $0.4\pm0.4$    &    $0.0^{+0.4}_{-0.4}$    &    \nodata    &    8.13e-04 \\

10360722    &   19 55 35.702&   +47 27 47.12&    15.03    &    0.725    &    5144    &    $-0.01$\phn\phs    &    4.66    &    $5124^{+199}_{-175}$    &    $4.13^{+0.27}_{-0.18}$    &    $1.03^{+0.21}_{-0.19}$    &    $1.47^{+0.52}_{-0.51}$    &    $1.45^{+2.06}_{-1.03}$    &    $0.5\pm0.5$    &    $-0.02^{+0.44}_{-0.45}$    &    \nodata    &    5.58e-04 \\

10525077 &   19 09 30.737&   +47 46 16.28&    15.36    &    0.536    &    5773    &    $-0.05$\phn\phs    &    4.42    & $    5736^{+     213}_{-     220}$ & $       3.50^{+       0.22}_{-       0.00}$ & $       1.16^{+       0.62}_{-       0.14}$ & $       3.00^{+       0.56}_{-       0.39}$ & $       3.4^{+      10.4}_{-       1.6}$ & $    0.06\pm     0.03$ & $      0.0^{+       0.4}_{-       0.4}$ & \nodata &    5.52e-04 \\

10850327    &   19 06 21.895&   +48 13 12.97&    13.01    &    0.382    &    6021    &    $-0.46$\phn\phs    &    4.44    &    $6022^{+181}_{-198}$    &    $4.1^{+0.2}_{-0.3}$    &    $1.04^{+0.20}_{-0.14}$    &    $1.40^{+1.04}_{-0.42}$    &    $1.97^{+3.09}_{-0.89}$    &    $0.5\pm0.8$    &    $-0.4^{+0.5}_{-0.5}$    &    \nodata    &    3.25e-04 \\

11026582    &   19 21 00.202&   +48 33 31.32&    14.47    &    0.450    &    5931    &    $-0.27$\phn\phs    &    4.60    &    $5934^{+183}_{-179}$    &    $4.2^{+0.1}_{-0.2}$    &    $1.04^{+0.14}_{-0.16}$    &    $1.19^{+0.49}_{-0.23}$    &    $1.62^{+1.18}_{-0.70}$    &    $0.9\pm0.8$    &    $-0.2^{+0.5}_{-0.4}$    &    \nodata    &    3.48e-04 \\

11253827$^{\dagger\dagger}$ &   19 44 31.877&   +48 58 38.64&    11.92    &    0.567    &    5384    &    $-0.13$\phn\phs    &    4.32    & $    5670^{+      86}_{-      81}$ & $       4.53^{+       0.04}_{-       0.05}$ & $       1.02^{+       0.03}_{-       0.06}$ & $       0.90^{+       0.06}_{-       0.05}$ & $       0.96^{+       0.15}_{-       0.15}$ & $      1.94\pm      0.39$ & $       0.17^{+       0.06}_{-       0.05}$ & $3.31(1.00)$ &    8.20e-03 \\

11392618$^{\dagger\dagger}$ &   19 03 37.454&   +49 17 15.07&    12.04    &    0.490    &    5520    &    $-0.47$\phn\phs    &    4.51    &$    5710^{+      39}_{-      43}$ & $       4.28^{+       0.04}_{-       0.04}$ & $       1.02^{+       0.03}_{-       0.03}$ & $       1.21^{+       0.07}_{-       0.06}$ & $       1.09^{+       0.24}_{-       0.15}$ & $     0.81\pm      0.14$ & $       0.01^{+       0.03}_{-       0.02}$ & $1.62(1.04)$  &    1.15e-04 \\

11716643    &   19 35 27.665&   +49 48 01.04&    14.69    &    0.551    &    5582    &    $-0.12$\phn\phs    &    4.74    &    $5565^{+191}_{-187}$    &    $4.5^{+0.2}_{-0.3}$    &    $0.96^{+0.13}_{-0.13}$    &    $0.85^{+0.63}_{-0.19}$    &    $0.83^{+0.82}_{-0.35}$    &    $2.2\pm3.2$    &    $-0.1^{+0.4}_{-0.4}$    &    \nodata    &    6.62e-04 \\
12735740$^{\dagger\dagger}$    $^{,\dagger\dagger\dagger}$ &   19 19 03.264&   +51 57 45.36&    12.62    &    0.478    &    5736    &    $-0.07$\phn\phs    &    4.34    &    $5629^{+42}_{-45}$    &    $4.408^{+0.044}_{-0.044}$    &    $0.94^{+0.02}_{-0.02}$    &    $1.00^{+0.05}_{-0.05}$    &    $0.79^{+0.09}_{-0.08}$    &    $1.3\pm0.3$    &    $-0.078^{+0.032}_{-0.028}$    &    1.43(0.78)    &    5.31e-04 \\

\enddata

\tablecomments{Stellar photometric variability is listed for each star in the last column as $\delta f/f$. Parameters marked with a $^{\dagger}$ are values from the KIC. Stars observed with Keck $HIRES$ are marked with a $^{\dagger\dagger}$. Stars observed with Keck NIRC2 are marked with a $^{\dagger\dagger \dagger}$. We assume $T_{\rm{eff}}=6000\pm500$ K, $\rm{[Fe/H]}=0.0\pm0.5$, and $\rm{log}\ g=4.5\pm0.5$ for KIC~8012732 because no KIC values are found. The assumption is based on $(g-r)$ value.}

\end{deluxetable}

\clearpage
 \end{landscape}

\newpage

\clearpage

\begin{landscape}
\begin{deluxetable}{lcccccccccccccc}
\tabletypesize{\tiny}
 \setlength{\tabcolsep}{0.02in} 
\tablewidth{0pt}
\tablecaption{Orbital Parameters\label{tab:orbital_params}}
\tablehead{
\colhead{\textbf{Star}} &
\colhead{\textbf{$T_0$}} &
\colhead{\textbf{Period}} &
\colhead{\textbf{Impact}} &
\colhead{\textbf{R$_{\rm PL}$/R$_{\ast}$}} &
\colhead{\textbf{e}} &
\colhead{\textbf{$\omega$}} &
\colhead{\textbf{R$_{\rm PL}$}} 	&
\colhead{\textbf{Inclination}} &
\colhead{\textbf{a/R$_{\ast}$}}	&
\colhead{\textbf{a}}	&
\colhead{\textbf{$T_{\rm PL}$}}	&
\colhead{\textbf{Depth}}	&
\colhead{\textbf{Dur}}	&
\colhead{\textbf{Sig}}	\\
\colhead{\textbf{(KIC)}} &
\colhead{\textbf{(MJD)}}	&
\colhead{\textbf{(d)}}	&
\colhead{\textbf{Parameter}}	&
\colhead{\textbf{ }}	&
\colhead{\textbf{ }}	&
\colhead{\textbf{(radian)}}	&
\colhead{\textbf{(R$_{\oplus}$)}}	&
\colhead{\textbf{($\deg$)}} &
\colhead{\textbf{}} &
\colhead{\textbf{(AU)}} &
\colhead{\textbf{(K)}} &
\colhead{\textbf{(ppm)}} &
\colhead{\textbf{(hr)}} &
\colhead{\textbf{($\sigma$)}}
}

\startdata

2581316 & 55071.3190 & $    217.8320\pm0.0063$ & $       0.245^{+       0.068}_{-       0.028}$ & $       0.0886^{+       0.0002}_{-       0.0002}$ & $       0.27^{+       0.09}_{-       0.27}$ & $       0.04^{+       0.36}_{-       0.04}$ & $ 20.31\pm  1.86$ & $      89.82^{+       0.02}_{-       0.08}$ & $      78.9^{+       2.4}_{-       8.4}$ & $      0.772\pm     0.067$ & $  459\pm   18$ & $9224\pm   82   $ &  21.4& 1.2 \\ 
 2975770    & 55130.8673  &$  369.0771\pm{0.0081 }$&$ 1.00  ^{ +0.00  }_{ -0.61 }$&$ 0.029  ^{ +0.021  }_{ -0.001 }$&$ 0.71  ^{+ 0.06  }_{ -0.47 }$&$ 0.45  ^{ +2.21  }_{ -0.45 }$&$ 2.61  \pm{ 3.03  }$&$ 89.72 ^{ +0.17  }_{ -0.98 }$&$ 246.3 ^{ +21.8  }_{ -156.4 }$&$ 0.961  \pm{ 0.082  }$&$ 205   \pm{ 40    }$&$ 1302  \pm{ 300   }$ & 6.0 & 0.1   \\
3326377    & 54989.0207  &$  198.7133\pm{0.0012 }$&$ 0.00  ^{+ 0.27  }_{ -0.00 }$&$ 0.043  ^{ +0.002  }_{ -0.001 }$&$ 0.00  ^{ +0.37  }_{ -0.00 }$&$ 0.00  ^{ +0.09  }_{ -0.00 }$&$ 2.97  \pm{ 0.23  }$&$ 90.00 ^{ +0.00  }_{ -0.08 }$&$ 195.5 ^{ +7.3   }_{ -9.4  }$&$ 0.584  \pm{ 0.016  }$&$ 240   \pm{ 13    }$&$ 2209  \pm{ 312   }$ & 8.2 & 0.6   \\
3634051    &    55192.5939    &    453.5264$\pm$0.0699    &    $0.811^{+0.066}_{-0.804}$    &    $0.0306^{+0.0024}_{-0.0042}$    &    $0.01^{+0.267}_{-0.006}$    &    $0.00^{+1.449}_{-0.00}$    &    7.36$\pm$2.58    &    $89.63^{+0.372}_{-0.124}$    &    $124.0^{+60.40}_{-22.16}$    &    1.205$\pm$0.082    &    353$\pm$63    &    898$\pm$162     &  18.6& 0.3 \\
3663173    &    55191.6724    &    174.6258$\pm$0.0077    &    $0.000^{+0.778}_{-0.000}$    &    $0.0476^{+0.0014}_{-0.0004}$    &    $0.28^{+0.234}_{-0.280}$    &    $0.65^{+0.989}_{-0.650}$    &    9.53$\pm$2.87    &    $90.00^{+0.000}_{-0.856}$    &    $71.2^{+14.63}_{-25.17}$    &    0.640$\pm$0.036    &    466$\pm$50    &    2928$\pm$965     &  15.7& 1.2 \\
3732035    &    54986.4187    &    138.9450$\pm$0.0256    &    $0.714^{+0.383}_{-0.433}$    &    $0.0195^{+0.0059}_{-0.0004}$    &    $0.35^{+0.043}_{-0.346}$    &    $1.57^{+0.179}_{-1.574}$    &    4.14$\pm$1.71    &    $89.27^{+0.541}_{-0.503}$    &    $56.0^{+31.94}_{-11.09}$    &    0.541$\pm$0.034    &    490$\pm$89    &    523$\pm$230     &  11.9& 0.6 \\
4142847    & 55133.2200  &$  210.6358\pm{0.0013 }$&$ 0.00  ^{ +0.61  }_{ -0.00 }$&$ 0.039  ^{ +0.007  }_{ -0.001 }$&$ 0.00  ^{+ 0.55  }_{ -0.00 }$&$ 0.00  ^{ +4.07  }_{ -0.00 }$&$ 2.85  \pm{ 0.41  }$&$ 90.00 ^{ +0.00  }_{ -0.21 }$&$ 197.9 ^{ +13.2  }_{ -14.7 }$&$ 0.613  \pm{ +0.023  }$&$ 226   \pm{ 25    }$&$ 2048  \pm{ 653   }$ & 8.4 & 0.3   \\
4472818 & 55086.1529 & $      70.9663 \pm0.0213$ & $       0.552^{+       0.121}_{-       0.154}$ & $       0.0251^{+       0.0016}_{-       0.0016}$ & $       0.00^{+       0.11}_{-      0.00}$ & $       0.00^{+       0.26}_{-       0.00}$ & $  3.18\pm  0.32$ & $      89.46^{+       0.18}_{-       0.20}$ & $      58.6^{+       6.4}_{-       7.1}$ & $      0.345\pm    0.008$ & $  506\pm   30$ & $803\pm   131    $ &   7.0& 0.7 \\
4760478$^{\dagger}$    &    55468.6953    &    
$  287.42680\pm0.00466  $&
$ 0.66  ^{ +0.27  }_{ -0.45 }$&
$ 0.057  ^{ +0.011  }_{ -0.007 }$&
$ 0.00  ^{ +0.26  }_{ -0.00 }$&
$ 0.00  ^{ +1.54  }_{ -0.00 }$&
$ 14.82 \pm{ 8.82  }$&
$ 89.55 ^{ +0.35  }_{ -0.72 }$&
$ 79.2  ^{ +20.6  }_{ -32.4 }$&
$ 0.894  \pm{ 0.108  }$&
$ 393   \pm{ 86    }$&
$ 3826  \pm{ 883   }$& 
20.0& 1.5 \\
4820550    &    55124.5479    &    202.1175$\pm$0.0009    &    $0.893^{+0.017}_{-0.032}$    &    $0.0609^{+0.0002}_{-0.0040}$    &    $0.00^{+0.04}_{-0.00}$    &    $0.00^{+0.16}_{-0.00}$    &    5.13$\pm$0.32    &    $89.72^{+0.02}_{-0.02}$    &    $189.2^{+9.2}_{-9.5}$    &    0.678$\pm$0.007    &    274$\pm$7    &    2801$\pm$200     &   4.6& 1.0 \\
4902202    & 55224.3199  &$  216.4667\pm{0.0057 }$&$ 0.00  ^{ +0.68  }_{ -0.00 }$&$ 0.032  ^{ +0.004  }_{ -0.004 }$&$ 0.01  ^{ +0.47  }_{ -0.01 }$&$ 0.00  ^{ +1.12  }_{ -0.00 }$&$ 2.89  \pm{ 2.23  }$&$ 90.00 ^{+ 0.00  }_{ -0.47 }$&$ 172.4 ^{ +12.7  }_{ -95.6 }$&$ 0.673  \pm{ 0.034  }$&$ 259   \pm{ 47    }$&$ 1319  \pm{ 705   }$  & 10.4 & 0.2   \\
4947556    &    55022.7737    &    141.6091$\pm$0.0045    &    $0.095^{+0.053}_{-0.095}$    &    $0.0324^{+0.0006}_{-0.0004}$    &    $0.00^{+0.01}_{-0.00}$    &    $0.00^{+0.04}_{-0.00}$    &    2.60$\pm$0.08    &    $89.96^{+0.03}_{-0.02}$    &    $149.8^{+4.1}_{-5.2}$    &    0.512$\pm$0.006    &    277$\pm$6    &    1306$\pm$195     &   6.9& 0.2 \\
5437945.01 & 55078.4788 & $     220.1344\pm0.0080$ & $       0.690^{+       0.476}_{-       0.144}$ & $       0.0407^{+       0.0009}_{-       0.0005}$ & $       0.34^{+       0.24}_{-       0.346}$ & $       0.22^{+       1.36}_{-       0.22}$ & $ 12.15\pm  1.95$ & $      89.33^{+       0.21}_{-       0.97}$ & $      59.5^{+       9.5}_{-      18.6}$ & $      0.748\pm     0.072$ & $  525\pm   64$ & $1804\pm   153    $ &  20.7& 1.6 \\
5437945.02 & 54971.8491 & $    440.7704\pm0.0015$ & $       0.530^{+       0.823}_{-       0.530}$ & $       0.0486^{+       0.0005}_{-       0.0017}$ & $       0.17^{+       0.53}_{-       0.17}$ & $       0.36^{+       1.22}_{-       0.36}$ & $  9.73\pm  3.22$ & $      89.77^{+       0.23}_{-       0.93}$ & $     134.4^{+      29.1}_{-      68.7}$ & $       1.18\pm     0.067$ & $  348\pm   64$ & $2646\pm   159    $ &  21.8& 1.7 \\
5857656$^{\dagger}$    &    55241.1260    &    
$  313.06282\pm0.00909  $&
$ 1.00  ^{ +0.00  }_{ -0.54 }$&
$ 0.024  ^{ +0.008  }_{ -0.001 }$&
$ 0.55  ^{ +0.21  }_{ -0.27 }$&
$ 1.51  ^{ +1.96  }_{ -1.42 }$&
$ 4.81  \pm{ 5.10  }$&
$ 89.29 ^{ +0.56  }_{ -1.19 }$&
$ 107.0 ^{ +39.1  }_{ -59.9 }$&
$ 0.948  \pm{ 0.132  }$&
$ 396   \pm{ 104   }$&
$ 630   \pm{ 107   }$  &
12.0& 0.5 \\
5871985    & 55313.8940  &$  213.2582\pm{0.0043 }$&$ 0.29  ^{ +0.39  }_{ -0.29 }$&$ 0.038  ^{ +0.004  }_{ -0.002 }$&$ 0.00  ^{+ 0.30  }_{ -0.00 }$&$ 0.00  ^{ +3.09  }_{ -0.00 }$&$ 2.39  \pm{ 0.28  }$&$ 89.92 ^{ +0.08  }_{ -0.21 }$&$ 219.2 ^{ +12.6  }_{ -13.9 }$&$ 0.584  \pm{ 0.018  }$&$ 201   \pm{ 26    }$&$ 1968  \pm{ 438   }$ & 7.6 & 0.8   \\
5966810    &    54966.4768    &    247.8919$\pm$0.0041    &    $0.000^{+0.690}_{-0.00}$    &    $0.0308^{+0.0014}_{-0.0011}$    &    $0.30^{+0.311}_{-0.299}$    &    $0.99^{+0.634}_{-0.986}$    &    6.81$\pm$1.81    &    $90.00^{+0.000}_{-0.563}$    &    $81.7^{+23.84}_{-20.65}$    &    0.804$\pm$0.051    &    440$\pm$49    &    1144$\pm$329     &  20.2& 0.4 \\
6106282    & 55048.0109  &$  101.1144\pm{0.0012}$&$ 0.00  ^{ +0.27  }_{ -0.00 }$&$ 0.032  ^{ +0.001  }_{ -0.001 }$&$ 0.00  ^{ +0.23  }_{ -0.00 }$&$ 0.00  ^{ +0.53  }_{ -0.00 }$&$ 2.04  \pm{ 0.23  }$&$ 90.00 ^{ +0.00  }_{ -0.13 }$&$ 132.4 ^{ +8.9   }_{ -10.4 }$&$ 0.359  \pm{ 0.013  }$&$ 236   \pm{ 21    }$&$ 1065  \pm{ 452   }$ & 6.4 & 0.4   \\
6878240    &    55066.1900    &    135.4985$\pm$0.0124    &    $0.796^{+0.108}_{-0.109}$    &    $0.0771^{+0.0029}_{-0.0086}$    &    $0.00^{+0.30}_{-0.00}$    &    $0.00^{+1.60}_{-0.00}$    &    8.46$\pm$1.43    &    $89.57^{+0.08}_{-0.14}$    &    $108.6^{+8.1}_{-18.3}$    &    0.493$\pm$0.021    &    326$\pm$23    &    5935$\pm$636     &   6.7& 5.3 \\
7826659    & 55111.6075  &$  211.0387\pm{0.0011 }$&$ 0.37  ^{ +0.31  }_{ -0.37 }$&$ 0.037  ^{ +0.004  }_{ -0.003 }$&$ 0.36  ^{ +0.19  }_{ -0.24 }$&$ 0.56  ^{ +2.29  }_{ -0.56 }$&$ 3.18  \pm{ 0.36  }$&$ 89.88 ^{ +0.12  }_{ -0.10 }$&$ 180.2 ^{ +7.3   }_{ -6.2  }$&$ 0.650  \pm{ 0.014  }$&$ 231   \pm{ 13    }$&$ 2009  \pm{ 194   }$ & 6.5 & 0.4   \\
8012732 & 55224.1613 & $     431.469\pm0.257$ & $       1.00^{+       0.00}_{-       0.27}$ & $       0.0741^{+       0.0007}_{-       0.0018}$ & $       0.44^{+       0.23}_{-       0.45}$ & $       0.60^{+       1.23}_{-       0.60}$ & $ 15.12\pm  5.03$ & $      89.37^{+       0.35}_{-       1.20}$ & $     107.7^{+      45.4}_{-      38.1}$ & $       1.18\pm     0.06$ & $  386\pm   79$ & $5513\pm   248    $ &  15.5& 1.4 \\
8210018    &    55226.3825    &    132.1730$\pm$0.00315    &    $0.536^{+0.395}_{-0.536}$    &    $0.0290^{+0.0068}_{-0.0007}$    &    $0.00^{+0.355}_{-0.000}$    &    $0.06^{+1.494}_{-0.059}$    &    2.88$\pm$1.35    &    $89.69^{+0.310}_{-0.560}$    &    $98.9^{+35.70}_{-37.69}$    &    0.434$\pm$0.029    &    285$\pm$45    &    1277$\pm$462     &   5.5& 0.3 \\
8636333$^{\dagger}$    &    55104.3899    &    804.7057$\pm$0.0107    &    $0.487^{+0.424}_{-0.461}$    &    $0.0450^{+0.0023}_{-0.0005}$    &    $0.36^{+0.316}_{-0.358}$    &    $1.07^{+0.542}_{-1.066}$    &    6.67$\pm$2.59    &    $89.89^{+0.106}_{-0.182}$    &    $251.2^{+102.1}_{-82.45}$    &    1.729$\pm$0.095    &    243$\pm$37    &    2396$\pm$515     &  18.8& 2.0 \\
8827930    &    55207.3096    &    288.3082$\pm$0.0074    &    $0.574^{+0.337}_{-0.574}$    &    $0.0992^{+0.0048}_{-0.0071}$    &    $0.15^{+0.269}_{-0.152}$    &    $0.57^{+1.034}_{-0.565}$    &    6.84$\pm$2.68    &    $89.90^{+0.100}_{-0.131}$    &    $326.6^{+82.04}_{-105.2}$    &    0.847$\pm$0.033    &    218$\pm$39    &    12711$\pm$782     &   5.7& 1.0 \\
9025971 & 55512.4290 & $    141.2412\pm0.0006$ & $       0.482^{+       0.827}_{-       0.194}$ & $       0.1086^{+       0.0007}_{-       0.0038}$ & $       0.25^{+       0.44}_{-       0.25}$ & $       0.00^{+       1.59}_{-       0.00}$ & $  9.59\pm  5.54$ & $      89.82^{+       0.08}_{-       0.86}$ & $     152.1^{+       9.9}_{-      80.7}$ & $      0.532 \pm     0.025$ & $  306\pm   46$ & $13712\pm   277 $ &   7.2& 0.9 \\
9147029    &    55251.2605    &    397.4995$\pm$0.0043    &    $0.577^{+0.222}_{-0.571}$    &    $0.0469^{+0.0042}_{-0.0011}$    &    $0.51^{+0.12}_{-0.50}$    &    $0.55^{+1.14}_{-0.55}$    &    4.65$\pm$0.35    &    $89.87^{+0.12}_{-0.05}$    &    $254.5^{+12.5}_{-11.3}$    &    1.075$\pm$0.012    &    244$\pm$6    &    2755$\pm$500     &   7.8& 0.7 \\
9166700    &    54967.4761    &    264.4510$\pm$0.046    &    $0.775^{+0.156}_{-0.756}$    &    $0.0220^{+0.0015}_{-0.0022}$    &    $0.01^{+0.460}_{-0.007}$    &    $0.00^{+1.554}_{-0.000}$    &    3.26$\pm$1.19    &    $89.78^{+0.216}_{-0.197}$    &    $201.2^{+86.17}_{-67.70}$    &    0.829$\pm$0.041    &    365$\pm$63    &    499$\pm$50     &   9.3& 0.5 \\
9413313$^{\dagger}$    &    55318.1058    &    441.0176$\pm$0.0036    &    $0.340^{+0.334}_{-0.279}$    &    $0.0769^{+0.0008}_{-0.0013}$    &    $0.14^{+0.498}_{-0.135}$    &    $0.37^{+1.232}_{-0.367}$    &    8.66$\pm$4.19    &    $89.94^{+0.051}_{-0.146}$    &    $316.4^{+31.72}_{-135.0}$    &    1.104$\pm$0.069    &    228$\pm$45    &    7435$\pm$255     &  10.0& 1.0 \\
9425139 & 55126.5676 & $     305.0620\pm0.0900$ & $       1.00^{+       0.00}_{-       0.60}$ & $       0.0633^{+       0.0003}_{-       0.0031}$ & $       0.46^{+       0.24}_{-       0.46}$ & $       0.83^{+       0.80}_{-       0.83}$ & $ 12.25\pm  4.99$ & $      89.52^{+       0.33}_{-       0.96}$ & $     164.8^{+     103.9}_{-      58.9}$ & $       0.92\pm      0.07$ & $  330\pm  111$ & $4007\pm   156    $ &   6.4& 0.2 \\
9480535    &    55035.9784    &    160.1288$\pm$0.0076    &    $0.712^{+0.074}_{-0.712}$    &    $0.0357^{+0.0001}_{-0.0070}$    &    $0.00^{+0.44}_{-0.00}$    &    $0.00^{+1.61}_{-0.00}$    &    7.44$\pm$2.26    &    $89.40^{+0.59}_{-0.20}$    &    $68.3^{+20.3}_{-15.3}$    &    0.606$\pm$0.033    &    469$\pm$63    &    1266$\pm$393     &  16.2& 0.3 \\
9663113$^{\dagger}$    &    55138.9999    &    572.3753$\pm$0.0141    &    $0.727^{+0.125}_{-0.069}$    &    $0.0429^{+0.0001}_{-0.0020}$    &    $0.00^{+0.258}_{-0.00}$    &    $0.00^{+1.497}_{-0.00}$    &    10.93$\pm$1.38    &    $89.69^{+0.051}_{-0.102}$    &    $136.1^{+11.74}_{-16.79}$    &    1.438$\pm$0.117    &    332$\pm$15    &    1790$\pm$186     &  23.0& 0.2 \\
9886255    &    55120.5100    &    105.9561$\pm$0.0041    &    $0.709^{+0.084}_{-0.539}$    &    $0.0510^{+0.0007}_{-0.0059}$    &    $0.00^{+0.28}_{-0.00}$    &    $0.52^{+1.00}_{-0.52}$    &    5.92$\pm$1.13    &    $89.53^{+0.36}_{-0.13}$    &    $87.2^{+14.3}_{-13.8}$    &    0.422$\pm$0.016    &    363$\pm$33    &    2601$\pm$569     &   8.5& 1.8 \\
9958387    &    55029.4656    &    237.7886$\pm$0.0053    &    $0.866^{+0.028}_{-0.020}$    &    $0.0479^{+0.0022}_{-0.0026}$    &    $0.00^{+0.09}_{-0.00}$    &    $0.00^{+0.00}_{-0.00}$    &    8.78$\pm$0.81    &    $89.51^{+0.04}_{-0.05}$    &    $102.4^{+4.7}_{-5.4}$    &    0.800$\pm$0.033    &    402$\pm$10    &    1671$\pm$147     &  10.4& 0.1 \\
10024862$^{\dagger}$    &    55192.1622    &    566.1537$\pm$0.0047    &    $0.758^{+0.145}_{-0.522}$    &    $0.0450^{+0.0003}_{-0.0030}$    &    $0.24^{+0.230}_{-0.243}$    &    $0.27^{+1.322}_{-0.271}$    &    7.51$\pm$2.17    &    $89.77^{+0.171}_{-0.160}$    &    $192.2^{+49.29}_{-56.86}$    &    1.388$\pm$0.079    &    305$\pm$39    &    2051$\pm$707     &  20.0& 0.1 \\
10360722    &    55071.2671    &    163.6915$\pm$0.0078    &    $0.963^{+0.039}_{-0.837}$    &    $0.0346^{+0.0056}_{-0.0038}$    &    $0.55^{+0.13}_{-0.52}$    &    $1.56^{+0.46}_{-1.55}$    &    5.56$\pm$2.11    &    $89.38^{+0.53}_{-0.31}$    &    $90.2^{+40.8}_{-21.7}$    &    0.593$\pm$0.038    &    358$\pm$63    &    1319$\pm$484     &   9.3& 0.9 \\
10525077$^{\dagger}$ & 55167.7516 & $     427.1250\pm0.8830$ & $       1.028^{+       0.452}_{-       0.257}$ & $       0.0566^{+       0.0078}_{-       0.0054}$ & $       0.195^{+       0.382}_{-       0.195}$ & $       0.99^{+       0.92}_{-       0.99}$ & $ 18.57\pm  3.67$ & $      89.1^{+       0.4}_{-       1.1}$ & $      64.7^{+      22.8}_{-      18.4}$ & $       1.17\pm      0.13$ & $  473\pm   77$ & $2803\pm   540    $ &  26.0& 0.2 \\
10850327$^{\dagger}$    &    55302.8600    &    440.1651$\pm$0.0500    &    $0.809^{+0.152}_{-0.378}$    &    $0.0279^{+0.0085}_{-0.0042}$    &    $0.16^{+0.093}_{-0.159}$    &    $0.54^{+0.673}_{-0.540}$    &    4.27$\pm$2.45    &    $89.76^{+0.154}_{-0.281}$    &    $189.6^{+78.17}_{-77.46}$    &    1.148$\pm$0.063    &    302$\pm$59    &    830$\pm$114     &  11.7& 1.4 \\
11026582    &    55071.9925    &    240.5778$\pm$0.0258    &    $0.000^{+0.659}_{-0.000}$    &    $0.0282^{+0.0014}_{-0.0003}$    &    $0.33^{+0.266}_{-0.325}$    &    $0.73^{+0.868}_{-0.732}$    &    3.69$\pm$1.13    &    $90.00^{+0.000}_{-0.318}$    &    $133.4^{+38.13}_{-33.99}$    &    0.768$\pm$0.038    &    337$\pm$43    &    970$\pm$236     &  10.5& 0.8 \\
11253827.01 & 54995.0098 & $      42.9910 \pm0.0186$ & $       0.391^{+       0.138}_{-       0.354}$ & $       0.0273^{+       0.0014}_{-       0.0007}$ & $       0.00^{+       0.05}_{-       0.00}$ & $       0.00^{+       0.03}_{-       0.00}$ & $  2.63\pm  0.24$ & $      89.62^{+       0.35}_{-       0.17}$ & $      58.4^{+       5.8}_{-       4.5}$ & $      0.242\pm    0.005$ & $  492\pm   22$ & $1078\pm   92    $ &   4.7& 0.2 \\
11253827.02 & 55103.6711 & $      88.5164 \pm0.0095$ & $       0.281^{+       0.217}_{-       0.282}$ & $       0.0447^{+       0.0022}_{-       0.0003}$ & $       0.00^{+       0.00}_{-       0.00}$ & $       0.00^{+       0.01}_{-       0.00}$ & $  4.42\pm  0.32$ & $      89.83^{+       0.17}_{-       0.16}$ & $      93.6^{+       5.8}_{-       8.1}$ & $      0.392\pm    0.006$ & $  389\pm   15$ & $2681\pm   103    $ &   7.1& 0.3 \\
11392618 & 55006.7218 & $     110.9232\pm0.0234$ & $       0.777^{+       0.0851}_{-       0.114}$ & $       0.0210^{+       0.0024}_{-       0.0042}$ & $       0.03^{+       0.11}_{-       0.03}$ & $       0.00^{+       1.33}_{-       0.00}$ & $  2.78\pm  0.47$ & $      89.45^{+       0.08}_{-       0.06}$ & $      81.7^{+       6.2}_{-       5.8}$ & $      0.456\pm    0.005$ & $  419\pm   15$ &$ 424\pm   89    $ &   5.9& 1.5 \\
11716643$^{\dagger}$    &    55267.4826    &    466.0661$\pm$0.0191    &    $0.010^{+0.787}_{-0.010}$    &    $0.0403^{+0.0076}_{-0.0004}$    &    $0.28^{+0.410}_{-0.282}$    &    $0.83^{+0.774}_{-0.830}$    &    3.75$\pm$1.87    &    $90.00^{+0.000}_{-0.202}$    &    $303.2^{+80.32}_{-123.7}$    &    1.162$\pm$0.055    &    218$\pm$38    &    2494$\pm$320     &  10.3& 0.6 \\
12735740    &    55195.5709    &    282.5255$\pm$0.0010    &    $0.512^{+0.050}_{-0.066}$    &    $0.0920^{+0.0002}_{-0.0004}$    &    $0.41^{+0.08}_{-0.29}$    &    $0.06^{+0.94}_{-0.06}$    &    10.12$\pm$0.56    &    $89.83^{+0.03}_{-0.02}$    &    $176.7^{+9.5}_{-9.0}$    &    0.828$\pm$0.009    &    281$\pm$7    &    9882$\pm$126     &  10.5& 0.6 \\

\enddata

\tablecomments{Pixel centroid offset significance is listed for each planet candidate in the last column. Planet candidates with only two transits are marked with a $^{\dagger}$.}

\end{deluxetable}

\clearpage
 \end{landscape}

\newpage

\clearpage

\begin{deluxetable}{lcc}
\tablewidth{0pt}
\tablecaption{Keck HIRES Doppler measurement results for KIC~12735740.\label{tab:RV_params}}
\tablehead{
\colhead{\textbf{Time}} &
\colhead{\textbf{RV}} &
\colhead{\textbf{uncertainty}} \\
\colhead{\textbf{(MJD)}} &
\colhead{\textbf{($\rm{m}\ \rm{s}^{-1}$)}} &
\colhead{\textbf{($\rm{m}\ \rm{s}^{-1}$)}} \\

}

\startdata

56446.894462 & 3.56 & 1.81 \\
56449.918787 & 18.26 & 1.72 \\
56451.065853 & -9.81 & 1.75 \\
56468.937706 & -12.00 & 2.18 \\

\enddata

\end{deluxetable}

\clearpage


\clearpage

\begin{deluxetable}{cc}
\tablewidth{0pt}
\tablecaption{Fraction of simulations in which a companion with a given mass is excluded at 3-$\sigma$ level based on Keck HIRES observations of KIC~12735740.\label{tab:simulation_params}}
\tablehead{
\colhead{\textbf{Mass}} &
\colhead{\textbf{Fraction}} \\
\colhead{\textbf{($M_J$)}} &
\colhead{\textbf{(\%)}} \\

}

\startdata

80 & 95.7 \\
40 & 91.0 \\
20 & 79.7 \\
10 & 55.5 \\

\enddata

\end{deluxetable}

\clearpage

\newpage

\end{document}